\documentclass{WileyMSP-template}
\usepackage{xcolor}
\definecolor{dark_green}{RGB}{0,120,0}
\usepackage{amsmath}
\usepackage{textcomp}

\usepackage{ragged2e}
\usepackage{setspace}
\usepackage[numbers,sort&compress]{natbib}

\newcommand{\abs}[1]{\left\lvert #1 \right\rvert}

\begin{document}

%\pagestyle{fancy}
%\rhead{\includegraphics[width=2.5cm]{vch-logo.png}}

\title{Sub-nm\textsuperscript{2} ferroelectric domains via charged 180° walls in ZrO\textsubscript{2}}

\maketitle

%Author: Please give full first and last names for authors and include * after the name of all corresponding authors

 \author{Nashrah Afroze\textsuperscript{1\dag}, Hamoon Fahrvandi\textsuperscript{2,3\dag}, Guodong Ren\textsuperscript{4\dag}, Pawan Kumar\textsuperscript{3}, Christopher Nelson\textsuperscript{5}, Sarah Lombardo\textsuperscript{6}, Mengkun Tian\textsuperscript{7}, Ping-Che Lee\textsuperscript{8}, Jiayi Chen\textsuperscript{1}, Manifa Noor\textsuperscript{9}, Kisung Chae\textsuperscript{8,9}, Sanghyun Kang\textsuperscript{1}, Prasanna Venkat Ravindran\textsuperscript{1}, Matthew Bergschneider\textsuperscript{9}, Gwan Yeong Jung\textsuperscript{10}, Pravan Omprakash\textsuperscript{4}, Gardy K. Ligonde\textsuperscript{11}, Nujhat Tasneem\textsuperscript{1}, Dina Triyoso\textsuperscript{12}, Steven Consiglio\textsuperscript{12}, Kanda Tapily\textsuperscript{12}, Robert Clark\textsuperscript{12}, Gert Leusink\textsuperscript{12}, Jayakanth Ravichandran\textsuperscript{13}, Shimeng Yu\textsuperscript{1}, Andrew Lupini\textsuperscript{5}, Andrew Kummel\textsuperscript{8}, Kyeongjae Cho\textsuperscript{9}, Duk-Hyun Choe\textsuperscript{14}, Nazanin Bassiri-Gharb\textsuperscript{11}, Josh Kacher\textsuperscript{6}, Rohan Mishra\textsuperscript{4,10*}, Jun Hee Lee\textsuperscript{3,15*}, Asif Khan\textsuperscript{1,6*}}

\begin{affiliations}

\textsuperscript{1}School of Electrical and Computer Engineering, Georgia Institute of Technology, Atlanta, GA, USA.\\

\textsuperscript{2}Department of Physics, Ulsan National Institute of Science and Technology, Ulsan, Republic of Korea.\\

\textsuperscript{3}Department of Energy Engineering, School of Energy and Chemical Engineering, Ulsan National Institute of Science and Technology, Ulsan, Republic of Korea.\\

\textsuperscript{4}Institute of Materials Science and Engineering, Washington University in St. Louis, St. Louis, MO, USA.\\

\textsuperscript{5}Center for Nanophase Materials Sciences, Oak Ridge National Laboratory, Oak Ridge, TN, USA.\\

\textsuperscript{6}School of Materials Science and Engineering, Georgia Institute of Technology, Atlanta, GA, USA.\\

\textsuperscript{7}Institute of Matter and Systems, Georgia Institute of Technology, Atlanta, GA, USA.\\

\textsuperscript{8}Department of Chemistry and Biochemistry, University of California San Diego, La Jolla, CA, USA.\\

\textsuperscript{9}Department of Materials Science and Engineering, University of Texas Dallas, Richardson, TX, USA.\\

\textsuperscript{10}Department of Mechanical Engineering and Material Science, Washington University in St. Louis, St. Louis, MO, USA.\\

\textsuperscript{11}School of Mechanical Engineering, Georgia Institute of Technology, Atlanta, GA, USA.\\

\textsuperscript{12}TEL Technology Center, America, LLC, Albany, NY USA.\\

\textsuperscript{13}Mork Family Department of Chemical Engineering and Materials Science, University of Southern California, Los Angeles, CA, USA.\\

\textsuperscript{14}Samsung Advanced Institute of Technology, Gyeonggi-do, Republic of Korea.\\
  
\textsuperscript{15}Graduate School of Semiconductor Materials and Devices Engineering, Ulsan National Institute of Science and Technology, Ulsan, Republic of Korea.\\

\textsuperscript{\dag} Nashrah Afroze, Hamoon Fahrvandi, and Guodong Ren contributed equally to this work. \\

\textsuperscript{*} Corresponding authors: Rohan Mishra, Jun Hee Lee, Asif Khan | Email: rmishra@wustl.edu, junhee@unist.ac.kr, akhan40@gatech.edu.\\
\end{affiliations}


\justifying
\doublespacing % Sets double spacing for the entire document

\begin{abstract}

Flat phonon bands in fluorite ferroelectrics (HfO$_2$ or ZrO$_2$) shrink polar domains laterally to an irreducible half–unit-cell width ($\sim$0.27 nm) within which the vertical arrangement of dipoles is expected to remain uniform. We report on the direct observation of nonuniform and nearly discrete vertical arrangements of dipoles in ZrO$_2$ thin films consisting of closely spaced head-to-head (HH) and tail-to-tail (TT) charged 180° walls, each exhibiting a distinct bulk-like structure. These charged domain walls (CDWs) further compress the irreducibly narrow, laterally stacked domains vertically to a thickness of $\sim$1–2.75 nm, yielding in-plane domains with sub-nm$^2$ footprints—among the smallest ever reported for any ferroelectric material. The HH and TT walls form due to their flat longitudinal optical (LO) polar bands and are electrostatically stabilized by bound-charge compensation via interstitial oxygen atoms, which act as natural structural defects at the HH walls. Moreover, these walls are predicted to be conducting and to exhibit ultralow propagation barriers, with HH walls (1.6 meV) being far more mobile than TT walls (22.3 meV), indicating strong potential for low-voltage, domain-wall-based nanoelectronics.

\end{abstract}

\section{Introduction}

Over the past decade, fluorite-structured binary oxide ferroelectrics—such as hafnium oxide (HfO$_2$), zirconium oxide (ZrO$_2$), and their alloys—have attracted interest for their seamless integration with complementary metal–oxide–semiconductor (CMOS) processes, combined with robust ferroelectricity down to the monolayer limit \cite{boscke2011ferroelectricity, cheema2022emergent}. At the heart of scalable ferroelectricity in this material class lies a remarkable phenomenon: anomalously flat, polar phonon bands drive the spontaneous formation of alternating lateral stacks of atomically narrow polar and nonpolar half–unit–cell layers. Consequently, polar distortions (i.e., dipoles) are laterally confined to irreducibly narrow two-dimensional (2D) domains separated by nonpolar spacers. These polar domains can be switched in a scale-free manner, enabled by flat-band-driven, neutral 180° domain walls \cite{lee2020scale}.

While the formation of neutral 180° domain walls in ferroelectric materials is facile, the stabilization of longitudinal, strongly charged 180° domain walls is generally energetically costly because of the significant depolarization fields arising from polarization bound charges and the large gradient of polar phonon bands involved in their formation \cite{kumar2025negative, gureev2011head, bednyakov2015formation}. Consequently, they are relatively rare in conventional ferroelectrics and typically appear as broad walls—spanning a few nanometers in PbTiO$_3$ \cite{gureev2011head} and Pb(Zr$_{0.2}$Ti$_{0.8})$O$_3$ \cite{jia2008atomic}, and tens of nanometers in BaTiO$_3$ \cite{bednyakov2015formation, gureev2011head}. In some multiferroics, such as hexagonal HoMnO$_3$, they can even extend up to hundreds of nanometers \cite{wu2012conduction}. When thinner charged 180° walls have been reported—for instance, $\sim$2 nm in PbTiO$_3$ \cite{moore2020highly}—they still appear as isolated single head-to-head (HH) or tail-to-tail (TT) walls, accompanied by disordered and widely extended polar domains spanning hundreds of nanometers. Stabilization of charged 180° walls may become even more challenging when such walls must form within extremely narrow, lateral polar domains, as in HfO$_2$ or ZrO$_2$. As a result, the vertical arrangement of dipoles within each domain has been expected to remain uniform \cite{wu2023unconventional, cheng2022reversible, choe2021unexpectedly, li2023polarization, zhou2022strain, park2024atomic, lee2021domains}.

In this article, using aberration-corrected scanning transmission electron microscopy (STEM) on ZrO$_2$ thin films, we report that within each laterally confined polar domains, dipoles can reverse their direction along the vertical axis without breaking the lateral coherence of the polar–nonpolar ordering. Thereby, dipoles meet alternately in HH and TT configurations, forming periodic, closely spaced, longitudinal charged 180° wall duplets, extending over only two unit cells, with $Pbcm$-like and $P4_2/nmc$-like core structures, respectively. Notably, our density functional theory (DFT) calculations reveal that the presence of relatively flat longitudinal optical (LO) polar bands at the HH and TT walls in ZrO$_2$ is key to their remarkable stabilization, close packing, and short spatial extent. From an electrostatic perspective, the polarization-induced charges at these charged domain walls (CDWs) are effectively screened by two interstitial oxygen atoms located at the Zr atomic layer at the center of each HH wall. Markedly, these CDWs restrict the laterally stacked domains vertically to a thickness of $\sim$1–2.75 nm. This, combined with the lateral confinement imposed by the pre-existing neutral 180° walls results in the stabilization of record-small in-plane ferroelectric domains with sub-nm$^2$ areas \cite{moore2020highly, li2025intrinsically, huyan2019structures}. Moreover, the HH and TT walls are predicted to be conducting and to exhibit ultralow motion barriers (1.6 meV for HH and 22.3 meV for TT), making them promising candidates for electric analogue of magnetic racetrack memories \cite{doi:10.1126/science.1154587,doi:10.1126/sciadv.1700512}.

\section{Results and Discussion}

\subsection{Experimental observation of head-to-head and tail-to-tail CDWs in ZrO$_2$}

We deposited a 10-nm ZrO$_2$ thin film on a Si substrate by atomic layer deposition, followed by annealing in N$_2$ at 350 $^\circ$C for 30 seconds (see Methods and Figure S1). Grazing incidence x-ray diffraction (GIXRD) shows dominance of orthorhombic $Pca2_1$ phase within the sample (Figures S2). Plan-view STEM imaging was performed in simultaneous high-angle annular dark-field (HAADF) and annular bright-field (ABF) modes (Figures S3 and S4). Figure 1a,b presents plan‐view ABF‐STEM micrographs of a selected region of the ZrO$_2$ film. Atomic mapping of the ABF‐STEM image confirms that this region is generally orthorhombic $Pca2_1$ viewed along [100], with columns of alternating polar and nonpolar half‐unit cells, laterally stacked along the [010] direction. While in the nonpolar layers the oxygen atoms (O\textsubscript{I}) are located at the centers of their respective Zr atomic cages, the oxygen atoms (O\textsubscript{II}) in the polar layers are displaced along the z direction, resulting in a net spontaneous polarization.

\begin{figure}[h!]
\includegraphics[width=\linewidth]{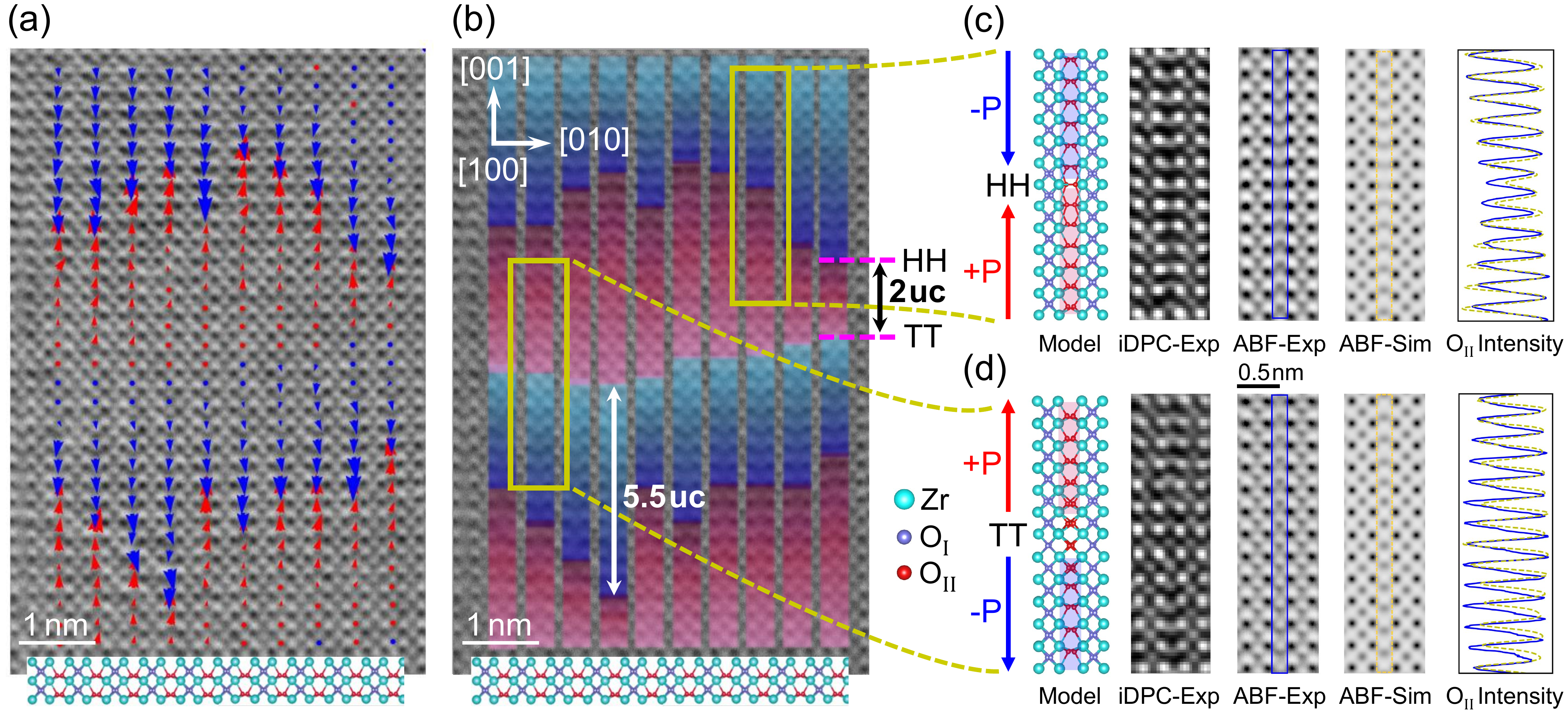} \caption{{\textbf{Observation of charged 180° domain-walls and antipolar ordering in ZrO$_2$.} \textbf{(a)} Plan-view ABF-STEM image of a [100]-oriented grain in a 10-nm-thick ZrO$_2$ film, revealing polar planes laterally separated by nonpolar layers. Each polar plane hosts a series of alternating, vertical HH and TT domains, separated by HH and TT charged 180° domain walls. The length of the dipole vectors represents the displacement of O\textsubscript{II} atoms from the centrosymmetric position of the respective surrounding Zr cage. The tetragonal $P4_2/nmc$ phase is regarded as the reference for determining the polarization direction. Note that atomic columns appear with dark contrast in the ABF-STEM image. \textbf{(b)} The corresponding down- and up-polarized domains, are denoted by blue and red colors, respectively}. The smallest and largest HH–TT domain-wall center-to-center distances are 2 and 5.5 unit cells (uc), respectively. For the \textbf{(c)} HH and \textbf{(d)} TT wall configurations, the atomic model, reconstructed atomic-potential map from DPC-STEM imaging, experimental ABF-STEM images, simulated ABF-STEM images, and oxygen-column intensity profiles are shown.}
\label{Fig1}
\end{figure}

Strikingly, the profile (Figure 1b) reveals that the dipole direction vertically reverses every 2–5.5 unit cells. Thereby, domains are not only confined laterally to a half-unit-cell width of $\sim$0.27 nm (along the [010] stacking axis) but also compressed vertically to $\sim$1–2.75 nm (corresponding to 2–5.5 unit cells), along the [001] polar axis, by charged 180° walls, resulting in the formation of the smallest in-plane ferroelectric domains with sub-nm$^2$ areas ($\sim$0.27–0.75 nm$^2$). To further confirm the presence of these ultra-small domains, we performed a template-matching analysis using a fast normalized cross-correlation method \cite{lewis1995fast}, as implemented in the scikit-image package \cite{van2014scikit} (Figure S5). The spatial variations of structural similarities with polar domains, assessed through template-matching, exhibit an alternating arrangement of HH and TT domains, consistent with the polar displacement vector measurements (Figure 1a).

The vertical alternation of HH and TT in-plane domains within each polar column gives rise to HH and TT CDWs. To elucidate the atomic structures of these CDWs, we present zoomed-in reconstructed atomic potential maps derived from DPC-STEM, and experimental ABF-STEM images (Figure 1c,d). Both the imaging techniques resolve oxygen columns alongside the heavier Zr atomic columns. At the HH wall core (Figure 1c), O\textsubscript{II} atoms shift into the Zr plane, resembling the nonpolar orthorhombic $Pbcm$ structure (Figure S6b). This $Pbcm$-like structure stabilizes the interface between two highly polar $Pca2_1$ domains of opposite polarization. At the TT wall core (Figure 1d), by contrast, the O\textsubscript{II} sublattice returns to near-center position within its Zr cage, resembling the $P4_2/nmc$ tetragonal structure (Figure S6c). STEM image simulations on DFT-optimized domain structures containing HH and TT walls using the multi-slice method, implemented in abtem \cite{Madsen2021abTEM} (Figure 1c,d), faithfully reproduce the ABF contrast, verifying our phase assignment. Intensity profiles of O\textsubscript{II} columns extracted from the simulated and experimental ABF images across the HH and TT walls further confirm the identified polar domain and domain wall configurations (right most panels of Figure 1c,d). We further corroborate the phase assignment at the HH and TT walls using phonon mode analysis, as discussed later. Across all the 38 domains imaged along the [100] zone axis (Figures S7–S10), we consistently observed alternating HH and TT motifs; the full set of O\textsubscript{II} displacement profiles for these walls is compiled in Figures S8 and S10. Moreover, the magnitudes of O\textsubscript{II} displacements are in good agreement with the simulated ABF image of a DFT-optimized structure (Figure S11). This confirms the reliability of the O\textsubscript{II} off-center displacements extracted from the ABF image. It is worth noting that, while multiple HH and TT wall structures were previously predicted by theory \cite{paul2024formation,xu2024oxygen,noor2024nearly,kumar2025negative}, their exact atomic configurations are experimentally observed here for the first time.

\subsection{Origin of the CDWs stabilization via first-principles simulations}

To elucidate the origin of the stabilization of the HH and TT CDWs (Figure 2a,b), we computed the phonon spectra of $Pbcm$ (Figure 2c) and $P4_2/nmc$ (Figure 2d) along the CDW direction ($\Gamma \rightarrow Z$) using DFT simulations. Both spectra reveal that the lowest-frequency longitudinal-optical (LO) phonons, $\Gamma_2^{\prime z}$ and $\Gamma_5^{\prime z}$—which reverse their signs during the formation of the HH and TT walls, respectively—exhibit remarkably narrow dispersions of approximately 30 cm$^{-1}$ with small band gradients along the Z-point wave vector. Since the band gradient energy ($E_{g}$), which contributes to the domain-wall energy, is proportional to the phonon band gradient \cite{kumar2025negative, shin2007nucleation}, these low-gradient, relatively flat LO bands significantly reduce the energy cost associated with CDWs formation. The band gradient energy is related to the band gradient according to the following equation:
\begin{equation}
E_{g} = \sum_{\lambda} m_{\lambda} g_{\lambda} \omega_{0\lambda} 
\sum_{q} u_{\lambda q}^{2} q^{2},
\end{equation}
%\begin{equation}
%E_{g} = \sum_{\lambda} m_{\lambda} g_{\lambda} \omega_{0\lambda} 
%\sum_{q} u_{\lambda q}^{2} q^{2} + O(m_{\lambda}, \omega_{0\lambda}, g_{\lambda}^{2}, q^{4}),
%\end{equation}
where $m_{\lambda}$, $g_{\lambda}$, and $\omega_{0\lambda}$ denote the effective mass, band gradient, and absolute frequency of the mode $\lambda$, respectively. The parameter $u_{\lambda q}$ indicates the mode’s reciprocal-space amplitude, and $q$ is the wave vector along the CDW direction. We estimated the gradient ($g_{\lambda}$) of each LO polar mode, by fitting its LO polar band along $\Gamma \rightarrow Z$, using the following quadratic equation:
\begin{equation}
\omega_{\lambda} = \omega_{0\lambda} + g_{\lambda} q^{2},
\end{equation}
where $\omega_{\lambda}$ is the frequency of mode $\lambda$ at any arbitrary wave vector $q$, and $\omega_{0\lambda}$ is the frequency of the LO polar modes ($\lambda = \Gamma_{2}^{\prime z}$ and $\Gamma_{5}^{\prime z}$) at the zone center in the HH and TT CDWs, respectively. We obtained remarkably small $g_{\lambda}$ of 42 and -85 cm$^{-1}$\,\AA$^{2}$ for the LO bands for the HH and TT walls, respectively. Such small $g_{\lambda}$ not only facilitates CDWs formation but also reduces their thickness ($\delta$), according to the relation $\delta \propto \sqrt{\abs{g_{\lambda}}}$ \cite{shin2007nucleation}. Moreover, our DFT-optimized ZrO$_2$ HH–TT wall supercell structure (Figure 2e) shows excellent agreement with our experimental observations, confirm the stabilization and coexistence of the observed periodic HH and TT CDWs, and corroborate their close packing and small thickness. Particularly, it shows a HH–TT center-to-center distance of only \(\sim\)1 nm, with HH and TT walls extending over \(\sim\)0.75 nm and \(\sim\)1 nm, respectively, compressing each laterally stacked, bulk-like polar domain vertically down to \(\sim\)0.25-0.5 nm. Moreover, the presence of the predicted $Pbcm$-like and $P4_2/nmc$-like phases at the HH and TT walls, respectively, is further verified through our DFT-based phonon-mode decomposition analysis (Figure S12b,c). The results demonstrate a convincing structural correspondence between these walls and their respective ideal bulk structures.

\begin{figure}[h!]
\centering
\includegraphics[width= 5.5in]{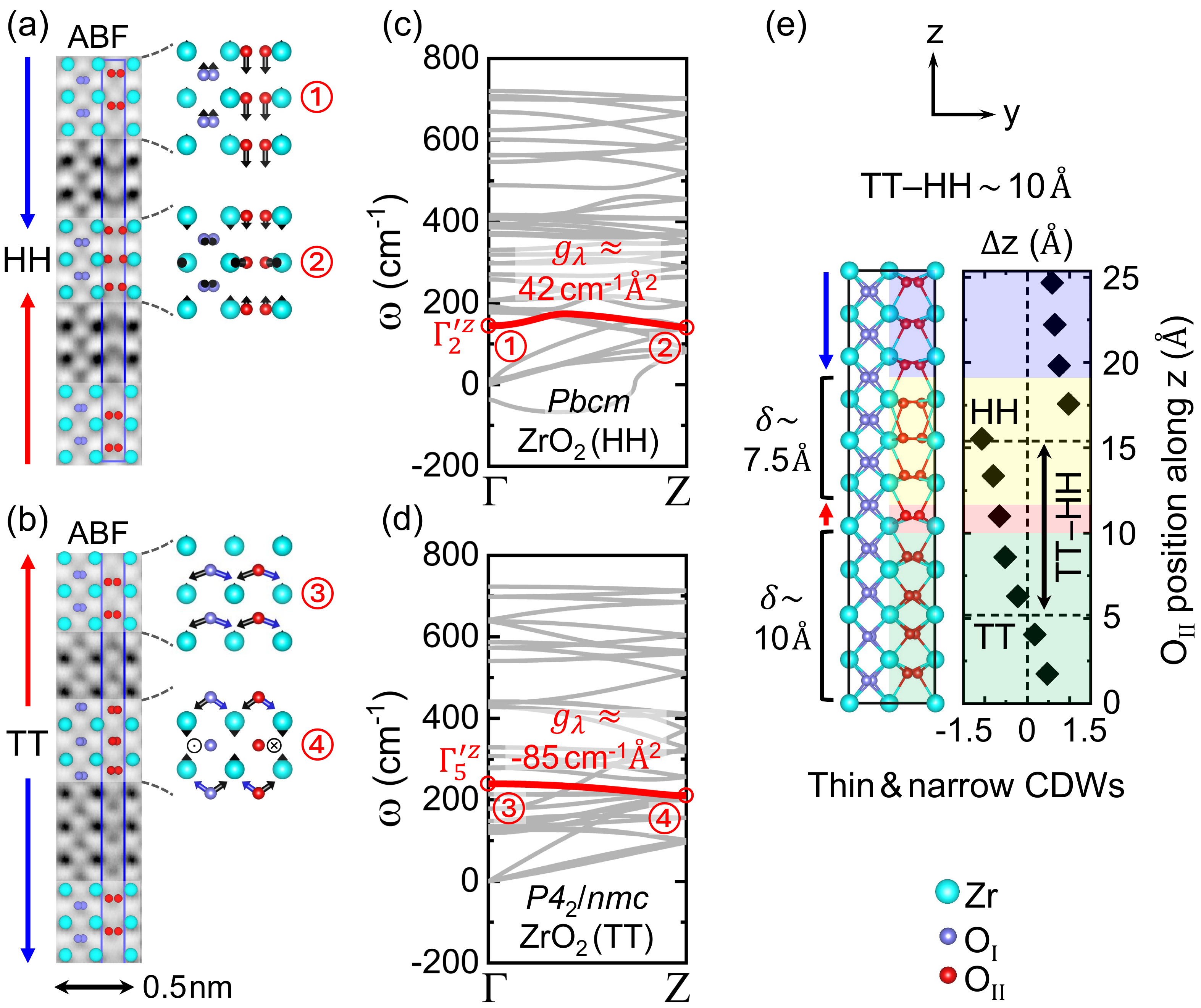}
\caption{{\textbf{Flat LO polar bands at the HH and TT domain walls in ZrO$_2$.} \textbf{(a)} and \textbf{(b)} Experimental ABF-STEM images for HH and TT domain wall configurations, featuring $Pbcm$- and $P4_2/nmc$-like core structures, respectively. Phonon band dispersions along the CDWs direction ($z$ direction) are plotted for \textbf{(c)} $Pbcm$ and \textbf{(d)} $P4_2/nmc$ phases of ZrO$_2$. The eigenvectors corresponding to the LO polar modes at $\Gamma$-point and their respective band-connected modes at $Z$-point are visualized (\textcircled{1}-\textcircled{4}) by black/blue arrows. The blue arrows denote the displacements of the back-plane atoms in \textcircled{3} and \textcircled{4}. The \textcircled{x} and \textcircled{$.$} show the displacements of oxygen atoms along –$x$ and $x$ directions, respectively.} \textbf{(e)} DFT-optimized ZrO$_2$ HH-TT configuration and the related off-center displacements of O\textsubscript{II} atoms.}
\label{Fig1}
\end{figure}

For comparison, we further calculated the phonon band dispersion for the cubic phase of perovskite PbTiO$_3$ (PTO), which represents its CDW core structure, along the $\Gamma \rightarrow Z$ direction. Our result exhibits a large LO bandwidth of approximately 167 cm$^{-1}$ with a considerable $g_{\lambda}$ of 386 cm$^{-1}$\,\AA$^{2}$ (Figure S13a). Consequently, the polar discontinuities in PTO CDWs incur significantly larger energy penalties and cannot remain sharply localized and closely spaced. Our DFT-optimized PTO HH–TT wall supercell structure (Figures S13b and S14) indicates HH–TT center-to-center distance of at least \(\sim\)6 nm, with each wall extending over \(\sim\)2.5 nm, leading to polar domains with the vertical thickness of \(\sim\)3.5 nm. We note that in addition to the significant LO band gradient difference between ZrO$_2$ and PTO, the lower frequencies of the Z-point modes (\textcircled{2} and \textcircled{4}) in ZrO$_2$, compared to the Z-point mode (\textcircled{6}) in PTO, indicate that the ZrO$_2$ lattice is intrinsically more compliant to the atomic rearrangements required at the interface. Thus, these two features—flat LO bands and relatively soft zone-boundary modes—drastically lower the formation energy of the HH and TT CDWs in ZrO$_2$, stabilizing them as duplets within each irreducibly narrow, laterally stacked domains.

To explain the origin of the LO-band flatness in ZrO$_2$, we analyzed its zone-boundary modes (\textcircled{2} and \textcircled{4}, Figures 2c,d), and found (i) the presence of a buffer atomic layer at the mid-plane of the cell along the $z$ direction, whose oxygen atoms show no polar displacements, thereby relatively isolating and localizing the opposite dipoles (or opposite polar domains) on the two sides of each HH/TT CDW, and (ii) pronounced  displacements of the heavy Zr atoms together with suppressed displacements of the light oxygen atoms (particularly along the $z$ direction), relative to those in the $\Gamma$-point modes (\textcircled{1} and \textcircled{3}, Figures 2c,d). While the first factor is the fundamental reason for the LO band flatness, the second factor may lower the zone-boundary mode's frequency, thereby fostering the flatness. In contrast, in PTO’s zone-boundary mode (\textcircled{6}, Figure S13a), these features are absent, giving rise to a higher mode frequency (251 cm$^{-1}$) compared to its $\Gamma$-point counterpart (83 cm$^{-1}$; \textcircled{5} in Figure S13a) and to a strongly dispersed LO band along the CDW direction.

We note that an additional crucial factor contributing to both the stabilization and short vertical spatial extent of the CDWs in ZrO$_2$ is the natural incorporation of two interstitial oxygen atoms (structural defects) at the center of each HH wall, which effectively suppress depolarization fields arising from polarization-bound charges (Figure S15). Interestingly, while the addition of these two oxygen atoms indirectly and largely compensates the bound charges at the TT wall through structural rearrangement, it substantially overcompensates the bound charges at the HH wall. This overcompensation arises because the two interstitial oxygen atoms contribute a total negative charge of 4e, whereas the initial (uncompensated) positive bound charge at each HH wall is only 1.63e, obtained as $2 \times P_s \times \text{Area}$.

\subsection{Electronic and dynamic characteristics of the CDWs}

\begin{figure}
\centering
\includegraphics[width= 6.5in]{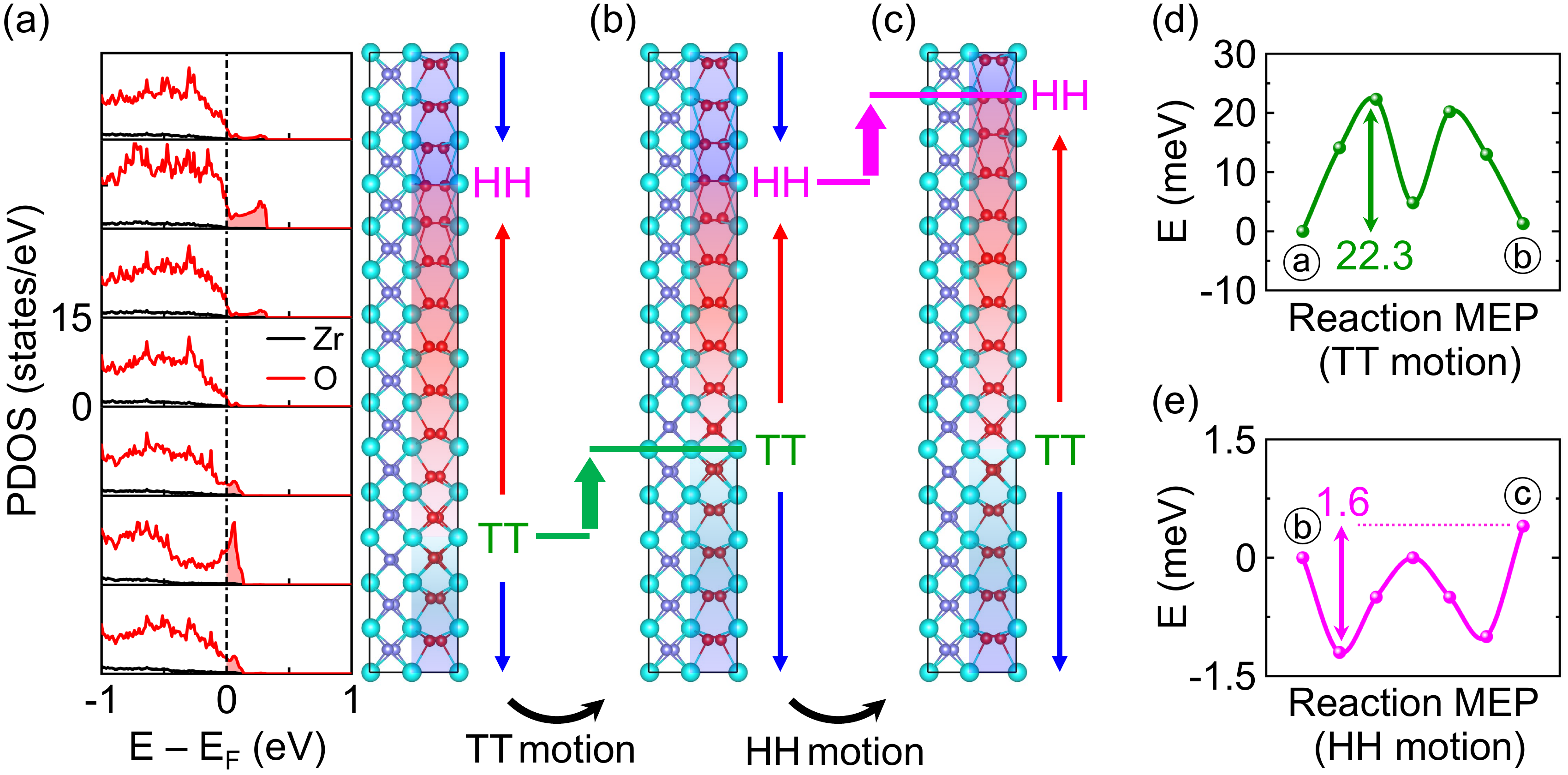}
\caption{\textbf{Projected density of states (PDOS) and HH/TT domain-wall dynamics.} \textbf{(a)} Unit-cell-by-unit-cell PDOS for a fully optimized HH–TT domain-wall system, highlighting hole-doped (p-type) metallic behavior at both TT and HH walls, particularly at the latter, due mainly to the interstitial oxygen atoms. The black and red lines show the PDOS for Zr and O atoms, respectively. \textbf{(b)} Translation of the TT domain wall by one unit cell with respect to the initial structure in panel (a), while the HH wall remains stationary. \textbf{(c)} Propagation of the HH wall with respect to the structure in panel (b), while the TT wall remains stationary. \textbf{(d)} and \textbf{(e)} The corresponding energy barriers and minimum-energy paths (MEP) for the motions of the TT and HH walls, respectively.}
\label{Fig1}
\end{figure}

To gain further insight into their broader implications for nanoscale memory and electronic applications, we investigated the electronic and dynamic properties of CDWs in ZrO$_2$. We calculated unit-cell-resolved partial densities of states (PDOS) for a fully optimized HH–TT CDW system (Figure 3a). Notably, the PDOS of the oxygen atoms reveals a pronounced metallic, p-type character at the HH wall, consistent with the overcompensation effect discussed above, arising from the presence of two interstitial oxygen atoms at this wall. At the TT wall, a less pronounced p-type state remains. In the domain/bulk regions, the electronic structure exhibits nearly ideal insulating behavior.

Furthermore, using the nudged elastic band (NEB) technique, we computed the minimum energy paths (MEPs) for the translation of TT and HH CDWs. Interestingly, the results reveal facile CDW propagation between polarization states (Figures 3a–c), with ultralow energy barriers of only 22.3 meV (Figure 3d) and 1.6 meV (Figure 3e) for TT and HH walls, respectively. We note that these walls can also propagate simultaneously, with a calculated barrier of 29.2 meV for this collective mode (Figure S16). Such remarkably low barriers, particularly for the HH wall, imply that these CDWs can be readily reconfigured and translated under minimal external stimuli. This property, combined with the conducting nature of these walls (Figure 3a), highlights their potential for ultrafast, energy-efficient domain-wall nanoelectronics and electric analogue of racetrack memories.

\subsection{Two-dimensional independence of the coexisting HH and TT domains}

\begin{figure}
\centering
\includegraphics[width= 6.5in]{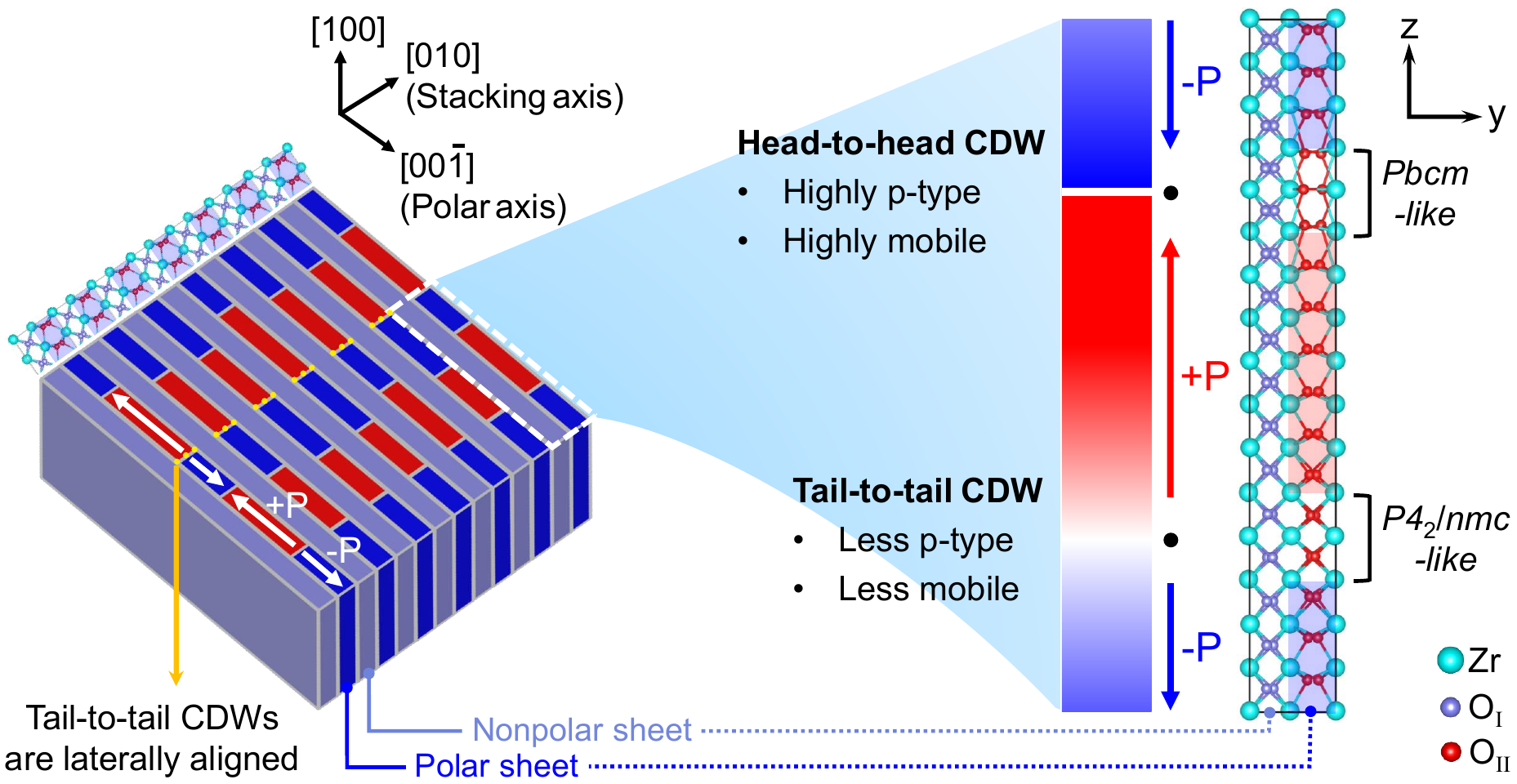}
\caption{{\textbf{Antipolar ordering \emph{via} strongly charged 180° domain walls.} Schematic plan‐view of a ZrO$_2$ film featuring self-organized, two-dimensional, irreducibly narrow, lateral polar layers (red-blue sheets) isolated by nonpolar planes (purple sheets). Each single polar layers contains a series of closely spaced, charged 180° domain walls alternating between HH and TT configurations, leading to vertical domains (red or blue) within each polar layer. The TT CDWs are laterally aligned in the adjacent polar sheets. The relaxed atomic model and schematic polarization profile for a selected region (dashed outline) of a polar sheet are shown on the right.}}
\label{Fig1}
\end{figure}

Our results show the coexistence of HH and TT CDWs, enabled by the relative flatness of the LO polar bands at both types of walls (the $Pbcm$ and $P4_2/nmc$ structures), together with bound-charge compensation effects. As a result, nanometer-scale vertical domains, sandwiched between alternating HH and TT CDWs, are formed in plane along the polar axis (Figure 4). Unlike the TT walls, HH walls need not align laterally across neighboring columns and can instead adopt independent positions along the polar axis. The lower motion barrier of HH walls compared with that of TT walls likely accounts for the independently random positioning of the former, whereas the latter remain nearly aligned under steady-state conditions.

Furthermore, the thickness of the antiparallel vertical domains varies both within individual 2D polar sheets and between adjacent sheets, revealing an incommensurate antipolar ordering. Lee et al.\cite{lee2020scale} demonstrated that the flat transverse-optical (TO) phonon band along the $\Gamma \rightarrow Y$ direction (i.e., the stacking axis) severs the perovskite-like three-dimensional elastic and electrostatic coupling between the adjacent dipolar planes, thereby effectively decoupling these planes from one another. Consequently, each dipolar sheet behaves as an independent polar “sandbox”, capable of sustaining distinct in-plane polarization patterns without incurring the depolarizing fields or elastic strains that would otherwise arise from misaligned walls in adjacent layers. Accordingly, each in-plane domain within a dipolar layer behaves as an independent stripe only half a unit cell wide along the [010] stacking axis. While the TO flat band isolates HH and TT domains laterally, the LO flat bands render them independent along the vertical direction. We therefore predict that, within each laterally stacked dipolar layer, the generated HH and TT CDWs not only propagate independently of neighboring walls within the same layer but also exhibit dynamics that are effectively isolated from those of adjacent dipolar layers.

\section{Conclusion}

Our results reveal a novel hierarchical self‐organization of dipoles in a ZrO$_2$ thin film: alternating polar—nonpolar ordering with single-unit-cell periodicity along the [010] stacking axis, and confined within each half-unit-cell-wide polar sheet, incommensurate antipolar ordered “domains” (at most a few unit cells long) enabled by ultrashort, strongly charged 180° HH and TT domain walls along the [001] polar axis, both driven by flat polar bands and bound-charge compensation. The coexistence of ferroelectric response, incommensurate antipolar ordering, and strongly charged domain walls in intimate 2D organization of laterally stacked, polar/nonpolar layers suggests an unprecedented polar malleability in these binary oxides—blurring the demarcations among ferroelectricity, antiferroelectricity, and hyperferroelectricity \cite{garrity2014hyperferroelectrics}—and presenting a new frontier for emergent phenomena in condensed matter physics. The extensive charged interfaces may host novel functionalities \cite{catalan2012domain, wu2012conduction, meier2012anisotropic, zhang2013direct} and facilitate a new pathway to stabilize negative capacitance for advanced transistor technologies \cite{kumar2025negative} and energy storage applications \cite{yang2025enhanced}. The antipolar ordering revealed here may underlie a novel mechanism for enabling antiferroelectricity in fluorite-structure oxides. The controlled generation, motion, and annihilation of complementary HH/TT walls under external stimuli \cite{zhong2026observation} could unlock new degrees of freedom to encode and manipulate bits at the near-atomic limit, ushering in a new paradigm for domain-wall nanoelectronics including racetrack memories with silicon-compatible simple oxides \cite{catalan2012domain,doi:10.1126/science.1154587,doi:10.1126/sciadv.1700512}.

\section{Methods}

%\begin{Methods}
%\section{Methods}
\subsection{Experimental methods}
\subsubsection{Device fabrication.}  
10 nm ZrO\textsubscript{2} was deposited by atomic layer deposition (ALD) in a 300 mm thin film formation tool, with the starting wafer being heavily doped p-type Si. The film deposition conditions are the same as previously described \cite{mukundan2019structural}. 12 nm thick top TiN electrodes were deposited on top of ZrO\textsubscript{2} using TiCl\textsubscript{4} and NH\textsubscript{3} plasma enhanced ALD. A post-deposition rapid thermal anneal in an N\textsubscript{2} atmosphere was performed at 350°C for 30 s to stabilize the ferroelectric properties of the ZrO\textsubscript{2}. Aluminum metal layers were e-beam evaporated to define the capacitor areas which also serve as a hard mask during subsequent wet etching (1:1 H\textsubscript{2}O\textsubscript{2}:H\textsubscript{2}O at 55 °C) of the top TiN electrode.

\subsubsection{Electrical measurements.}
Electrical characterization was performed using an aixACCT TF Analyzer 3000 ferroelectric parameter analyzer at 1 kHz. The top electrodes were removed via wet-etching with (1:1 H\textsubscript{2}O\textsubscript{2}:H\textsubscript{2}O at 55°C) and dipping in (20:1 H\textsubscript{2}O:HF) solution for 30s at room temperature after electrical characterization to isolate the ZrO\textsubscript{2} layer in preparation for GIXRD and subsequent plan-view imaging.

\subsubsection{GIXRD measurements.}  
Grazing-incidence X-ray diffraction (GIXRD) was performed using a RIGAKU Smartlab XE diffractometer equipped with a Cu K\(\alpha\) source (40kV, 50mA) and a HyPix-3000HE detector. Data was acquired in the range of 20°-55°, with an incidence angle of 0.5°, scanning step of 0.04°, and scanning speed of 0.05°/min.

\subsubsection{Cross-sectional sample preparation.} 
Cross-sectional samples for (S)TEM imaging were prepared using an FEI Nova Nanolab 200 FIB/SEM equipped with a high energy Focused Ion Beam (FIB) using Gallium-69 and operated at an accelerating voltage between 5 and 30 kV. The resulting FIB lamellas were approximately 20 \(\mu\)m in length with 5\(\mu\)m windows thinned to approximately 50 nm in thickness.

\subsubsection{Plan-view sample preparation.} 
After completing electrical and GIXRD measurements, a plan-view TEM sample was prepared from the bulk by first manually polishing the silicon side using a steel polishing stub. This was followed by thinning with a Gatan Model 656 Dimple Grinder. Final thinning to electron transparency was achieved by ion milling the sample from the silicon side using a Gatan Model 691 Precision Ion Polishing System (PIPS) (Figure S3).

\subsubsection{TEM, STEM and 4D-STEM imaging.} 
Transmission electron microscopy (TEM) images (Figure S1b) were collected using FEI Tecnai G2 F30 TEM/STEM operated at an accelerating voltage of 300 kV and a beam current of about 80–100 pA. Annular bright field (ABF), high angle annular dark field (HAADF) and 4D-STEM based DPC images were collected using 100KV accelerating voltage at 3rd generation C3/C5 aberration corrected Nion UltraSTEM 100 (U100) microscope at Oak Ridge National Laboratory (ORNL). The STEM employed a 20 pA probe current in probe size of \(\sim\)1 {\AA} and convergent semi-angle of 30mrad. The STEM-HAADF images were acquired using an annular dark-field image detector with an inner collection semi-angle of 80 mrad. The STEM-ABF images were acquired using a bright-field image detector with collection semi-angle range of 15 to 30 mrad. 4D-STEM based DPC datasets were collected with a convergence angle of 30 mrad with 256 × 256-pixel size using a Nion 2020 Ronchigram camera, with a Hamamatsu ORCA ultra-low noise scientific CMOS sensor. At each pixel position, a 64 × 64-pixel diffraction image was recorded with a dwell time of 2 ms. DPC-STEM imaging measures the center-of-mass shifts of the convergent beam electron diffraction (CBED) disks, reflects relative probe deflection induced by the local electric field and provides enhanced sensitivity to light elements \cite{hachtel2018sub}. A Hitachi HD-2700 aberration- corrected STEM/SEM was used to capture low mag STEM images at Georgia Tech (Figure S4a), operated at a 200 kV accelerating voltage and 30 mrad convergence semi-angle. ABF-STEM images of Fig. 1C-D were denoised using Gaussian filter with a sigma of 3 pixels. Band-pass Fourier filtering in Digital Micrograph software was applied to all the other ABF images of the manuscript to reduce noise and enhance atomic sharpness.

\subsubsection{STEM-ABF image analysis}
To determine the atomic positions of both Zr and O atoms from STEM-ABF images, the Atomap \cite{nord2017atomap} library in Python was utilized. Atomap employs a two-dimensional (2D) Gaussian fitting algorithm to accurately identify atomic positions. Before applying Atomap, the contrast of ABF images was inverted using inverted contrast extension in Digital Micrograph software to enhance the atomic contrast. The Atomap library was then employed in a three-step process: first, the positions of Zr atoms were identified, followed by the extraction of O atom positions. The displacement of O\textsubscript{II} atoms from the centrosymmetric position of the surrounding Zr cage was then calculated to quantify the magnitude of the polarization vectors. The tetragonal phase was chosen as the reference state for determining the polarization direction.       

\subsection{Multi-slice simulation}
Multi-slice simulation of 10nm HZO HH and TT slab structures were implemented in \textit{abtem} \cite{Madsen2021abTEM}. The defocus value was set to 10~\AA~to obtain good agreement with the experimental conditions. Simulations were performed using an aberration-free probe with an accelerating voltage of 100 kV and a convergence semi-angle of 30 mrad. DFT-optimized domain structures containing head-to-head (HH) and tail-to-tail (TT) charged 180° domain walls shown in Figures 1C,D were used for the STEM simulations.

%\subsection{NBED.} 
%NBED scans were collected using a FEI Tecnai G2 F30 TEM/STEM operated at an accelerating voltage of 300 kV and beam current of about 80-100 pA.

\subsection{Computational methods} 
First-principles calculations were carried out using the Vienna Ab initio Simulation Package (VASP) \cite{kresse1996efficient, kresse1996efficiency, kresse1993ab}, based on Projector Augmented Wave (PAW) \cite{blochl1994projector} method and a plane-wave basis set. The exchange-correlation functional was calculated through the generalized-gradient approximation (GGA) as formulated by Perdew-Burke-Ernzerhof for solids \cite{perdew2008restoring}. The plane-wave basis set of the wave functions was expanded up to a kinetic energy cutoff of 500 eV. For sampling the integrations over the Brillouin zone, we considered Monkhorst-Pack (MP) \cite{monkhorst1976special} k-point grids of \( 8 \times 8 \times 8 \) and \( 10 \times 10 \times 10 \) for ZrO\textsubscript{2} and PTO, respectively. We calculated the minimum energy path for CDW motion using the nudged elastic band (NEB) method \cite{henkelman2000improved}. The density functional perturbation theory (DFPT) approach was employed to investigate the phononic properties of the structures. We computed the phonon band dispersions along the Z-wavevector using \( 1 \times 1 \times 2 \) supercells. Non-analytical term correction was applied to dynamical matrix based on Gonze scheme \cite{gonze1994interatomic}, using PHONOPY code \cite{togo2015first}.

\section*{Acknowledgment} %delete if not applicable))
The work was supported by the National Science Foundation CAREER Award (award \# 2047880), ASCENT, one of six centers in the Joint University Microelectronics Program (JUMP) program, and SUPREME, one of the seven centers of JUMP 2.0. Both are Semiconductor Research Corporation (SRC) programs sponsored by the Defense Advanced Research Program Agency (DARPA). Sample fabrication and imaging, in part, was done at Georgia Tech Institute for matter and systems (IMS) which is supported by the National Science Foundation. STEM experiments at Oak Ridge National Lab were conducted as part of a user project at the Center for Nanophase Materials Sciences (CNMS), which is a US Department of Energy, Office of Science User Facility at Oak Ridge National Laboratory. The ab initio DFT computational work---conducted at Ulsan National Institute of Science and Technology (UNIST)---was supported by the Next-generation Intelligence Semiconductor R\&D Program (2022M3F3A2A01079710), Midcareer Researcher (2020R1A2C2103126), Basic Research Laboratory (RS-2023-00218799), Nano \& Material Technology Development Program (RS-2024-00404361), and RS-2023-00257666, through the National Research Foundation of Korea (NRF) funded by the Korean government (MSIT). Additional support for the computational work was provided by the Korea Institute for Advancement of Technology (KIAT) grant funded by the Korean Government (MOTIE) (P0023703, HRD Program for Industrial Innovation), and by the National Supercomputing Center, which provided supercomputing resources and technical support (KSC-2022-CRE-0075, KSC-2022-CRE-0454, KSC-2022-CRE-0456, KSC-2023-CRE-0547, KSC-2024-CRE-0545). The work at Washington University was supported by the National Science Foundation under grant numbers DMR-2145797 (G.R. \& P.O.) and DMR-2122070 (G.Y.J. \& R.M.). M.N., M.B., A.K., and K.C. are funded by Semiconductor Research Corporation (SRC), grant number 3146.001. Interpretation of the data (G.K.L. \& N.B.-G.) was in part supported by the National Science Foundation grant DMR-2026976.

\section*{Conflicts of Interest} 
The authors declare no conflicts of interest.

\medskip

%\bibliographystyle{naturemag}
%\bibliography{references}

\begin{thebibliography}{10}
\expandafter\ifx\csname url\endcsname\relax
  \def\url#1{\texttt{#1}}\fi
\expandafter\ifx\csname urlprefix\endcsname\relax\def\urlprefix{URL }\fi
\providecommand{\bibinfo}[2]{#2}
\providecommand{\eprint}[2][]{\url{#2}}

\bibitem{boscke2011ferroelectricity}
\bibinfo{author}{B{\"o}scke, T.}, \bibinfo{author}{M{\"u}ller, J.}, \bibinfo{author}{Br{\"a}uhaus, D.}, \bibinfo{author}{Schr{\"o}der, U.} \& \bibinfo{author}{B{\"o}ttger, U.}
\newblock \bibinfo{title}{Ferroelectricity in hafnium oxide thin films}.
\newblock \emph{\bibinfo{journal}{Applied Physics Letters}} \textbf{\bibinfo{volume}{99}} (\bibinfo{year}{2011}).

\bibitem{cheema2022emergent}
\bibinfo{author}{Cheema, S.~S.} \emph{et~al.}
\newblock \bibinfo{title}{Emergent ferroelectricity in subnanometer binary oxide films on silicon}.
\newblock \emph{\bibinfo{journal}{Science}} \textbf{\bibinfo{volume}{376}}, \bibinfo{pages}{648--652} (\bibinfo{year}{2022}).

\bibitem{lee2020scale}
\bibinfo{author}{Lee, H.-J.} \emph{et~al.}
\newblock \bibinfo{title}{Scale-free ferroelectricity induced by flat phonon bands in hfo2}.
\newblock \emph{\bibinfo{journal}{Science}} \textbf{\bibinfo{volume}{369}}, \bibinfo{pages}{1343--1347} (\bibinfo{year}{2020}).

\bibitem{kumar2025negative}
\bibinfo{author}{Kumar, P.}, \bibinfo{author}{Gupta, D.} \& \bibinfo{author}{Lee, J.~H.}
\newblock \bibinfo{title}{Negative gradient energy facilitates charged domain walls in hfo 2}.
\newblock \emph{\bibinfo{journal}{Physical Review Letters}} \textbf{\bibinfo{volume}{134}}, \bibinfo{pages}{166101} (\bibinfo{year}{2025}).

\bibitem{gureev2011head}
\bibinfo{author}{Gureev, M.~Y.}, \bibinfo{author}{Tagantsev, A.~K.} \& \bibinfo{author}{Setter, N.}
\newblock \bibinfo{title}{Head-to-head and tail-to-tail 180 domain walls in an isolated ferroelectric}.
\newblock \emph{\bibinfo{journal}{Physical Review B—Condensed Matter and Materials Physics}} \textbf{\bibinfo{volume}{83}}, \bibinfo{pages}{184104} (\bibinfo{year}{2011}).

\bibitem{bednyakov2015formation}
\bibinfo{author}{Bednyakov, P.~S.}, \bibinfo{author}{Sluka, T.}, \bibinfo{author}{Tagantsev, A.~K.}, \bibinfo{author}{Damjanovic, D.} \& \bibinfo{author}{Setter, N.}
\newblock \bibinfo{title}{Formation of charged ferroelectric domain walls with controlled periodicity}.
\newblock \emph{\bibinfo{journal}{Scientific reports}} \textbf{\bibinfo{volume}{5}}, \bibinfo{pages}{15819} (\bibinfo{year}{2015}).

\bibitem{jia2008atomic}
\bibinfo{author}{Jia, C.-L.} \emph{et~al.}
\newblock \bibinfo{title}{Atomic-scale study of electric dipoles near charged and uncharged domain walls in ferroelectric films}.
\newblock \emph{\bibinfo{journal}{Nature materials}} \textbf{\bibinfo{volume}{7}}, \bibinfo{pages}{57--61} (\bibinfo{year}{2008}).

\bibitem{wu2012conduction}
\bibinfo{author}{Wu, W.}, \bibinfo{author}{Horibe, Y.}, \bibinfo{author}{Lee, N.}, \bibinfo{author}{Cheong, S.-W.} \& \bibinfo{author}{Guest, J.}
\newblock \bibinfo{title}{Conduction of topologically protected charged ferroelectric domain walls}.
\newblock \emph{\bibinfo{journal}{Physical review letters}} \textbf{\bibinfo{volume}{108}}, \bibinfo{pages}{077203} (\bibinfo{year}{2012}).

\bibitem{moore2020highly}
\bibinfo{author}{Moore, K.} \emph{et~al.}
\newblock \bibinfo{title}{Highly charged 180 degree head-to-head domain walls in lead titanate}.
\newblock \emph{\bibinfo{journal}{Communications Physics}} \textbf{\bibinfo{volume}{3}}, \bibinfo{pages}{231} (\bibinfo{year}{2020}).

\bibitem{wu2023unconventional}
\bibinfo{author}{Wu, Y.} \emph{et~al.}
\newblock \bibinfo{title}{Unconventional polarization-switching mechanism in (hf, zr) o 2 ferroelectrics and its implications}.
\newblock \emph{\bibinfo{journal}{Physical Review Letters}} \textbf{\bibinfo{volume}{131}}, \bibinfo{pages}{226802} (\bibinfo{year}{2023}).

\bibitem{cheng2022reversible}
\bibinfo{author}{Cheng, Y.} \emph{et~al.}
\newblock \bibinfo{title}{Reversible transition between the polar and antipolar phases and its implications for wake-up and fatigue in hfo2-based ferroelectric thin film}.
\newblock \emph{\bibinfo{journal}{Nature communications}} \textbf{\bibinfo{volume}{13}}, \bibinfo{pages}{645} (\bibinfo{year}{2022}).

\bibitem{choe2021unexpectedly}
\bibinfo{author}{Choe, D.-H.} \emph{et~al.}
\newblock \bibinfo{title}{Unexpectedly low barrier of ferroelectric switching in hfo2 via topological domain walls}.
\newblock \emph{\bibinfo{journal}{Materials Today}} \textbf{\bibinfo{volume}{50}}, \bibinfo{pages}{8--15} (\bibinfo{year}{2021}).

\bibitem{li2023polarization}
\bibinfo{author}{Li, X.} \emph{et~al.}
\newblock \bibinfo{title}{Polarization switching and correlated phase transitions in fluorite-structure zro2 nanocrystals}.
\newblock \emph{\bibinfo{journal}{Advanced Materials}} \bibinfo{pages}{2207736} (\bibinfo{year}{2023}).

\bibitem{zhou2022strain}
\bibinfo{author}{Zhou, S.}, \bibinfo{author}{Zhang, J.} \& \bibinfo{author}{Rappe, A.~M.}
\newblock \bibinfo{title}{Strain-induced antipolar phase in hafnia stabilizes robust thin-film ferroelectricity}.
\newblock \emph{\bibinfo{journal}{Science Advances}} \textbf{\bibinfo{volume}{8}}, \bibinfo{pages}{eadd5953} (\bibinfo{year}{2022}).

\bibitem{park2024atomic}
\bibinfo{author}{Park, K.} \emph{et~al.}
\newblock \bibinfo{title}{Atomic-scale scanning of domain network in the ferroelectric hfo2 thin film}.
\newblock \emph{\bibinfo{journal}{ACS nano}} \textbf{\bibinfo{volume}{18}}, \bibinfo{pages}{26315--26326} (\bibinfo{year}{2024}).

\bibitem{lee2021domains}
\bibinfo{author}{Lee, D.~H.} \emph{et~al.}
\newblock \bibinfo{title}{Domains and domain dynamics in fluorite-structured ferroelectrics}.
\newblock \emph{\bibinfo{journal}{Applied Physics Reviews}} \textbf{\bibinfo{volume}{8}} (\bibinfo{year}{2021}).

\bibitem{li2025intrinsically}
\bibinfo{author}{Li, X.} \emph{et~al.}
\newblock \bibinfo{title}{Intrinsically stable charged domain walls in molecular ferroelectric thin films}.
\newblock \emph{\bibinfo{journal}{Advanced Electronic Materials}} \textbf{\bibinfo{volume}{11}}, \bibinfo{pages}{2400324} (\bibinfo{year}{2025}).

\bibitem{huyan2019structures}
\bibinfo{author}{Huyan, H.}, \bibinfo{author}{Li, L.}, \bibinfo{author}{Addiego, C.}, \bibinfo{author}{Gao, W.} \& \bibinfo{author}{Pan, X.}
\newblock \bibinfo{title}{Structures and electronic properties of domain walls in bifeo3 thin films}.
\newblock \emph{\bibinfo{journal}{National Science Review}} \textbf{\bibinfo{volume}{6}}, \bibinfo{pages}{669--683} (\bibinfo{year}{2019}).

\bibitem{doi:10.1126/science.1154587}
\bibinfo{author}{Hayashi, M.}, \bibinfo{author}{Thomas, L.}, \bibinfo{author}{Moriya, R.}, \bibinfo{author}{Rettner, C.} \& \bibinfo{author}{Parkin, S. S.~P.}
\newblock \bibinfo{title}{Current-controlled magnetic domain-wall nanowire shift register}.
\newblock \emph{\bibinfo{journal}{Science}} \textbf{\bibinfo{volume}{320}}, \bibinfo{pages}{209--211} (\bibinfo{year}{2008}).

\bibitem{doi:10.1126/sciadv.1700512}
\bibinfo{author}{Sharma, P.} \emph{et~al.}
\newblock \bibinfo{title}{Nonvolatile ferroelectric domain wall memory}.
\newblock \emph{\bibinfo{journal}{Science Advances}} \textbf{\bibinfo{volume}{3}}, \bibinfo{pages}{e1700512} (\bibinfo{year}{2017}).

\bibitem{lewis1995fast}
\bibinfo{author}{Lewis, J.~P.}
\newblock \bibinfo{title}{Fast normalized cross-correlation}.
\newblock In \emph{\bibinfo{booktitle}{Vision interface}}, vol.~\bibinfo{volume}{10}, \bibinfo{pages}{120--123} (\bibinfo{year}{1995}).

\bibitem{van2014scikit}
\bibinfo{author}{Van~der Walt, S.} \emph{et~al.}
\newblock \bibinfo{title}{scikit-image: image processing in python}.
\newblock \emph{\bibinfo{journal}{PeerJ}} \textbf{\bibinfo{volume}{2}}, \bibinfo{pages}{e453} (\bibinfo{year}{2014}).

\bibitem{Madsen2021abTEM}
\bibinfo{author}{Madsen, J.} \& \bibinfo{author}{Susi, T.}
\newblock \bibinfo{title}{The abtem code: Transmission electron microscopy from first principles}.
\newblock \emph{\bibinfo{journal}{Open Research Europe}} \textbf{\bibinfo{volume}{1}}, \bibinfo{pages}{24} (\bibinfo{year}{2021}).

\bibitem{paul2024formation}
\bibinfo{author}{Paul, T.~K.}, \bibinfo{author}{Saha, A.~K.} \& \bibinfo{author}{Gupta, S.~K.}
\newblock \bibinfo{title}{Formation and energetics of head-to-head and tail-to-tail domain walls in hafnium zirconium oxide}.
\newblock \emph{\bibinfo{journal}{Scientific Reports}} \textbf{\bibinfo{volume}{14}}, \bibinfo{pages}{9861} (\bibinfo{year}{2024}).

\bibitem{xu2024oxygen}
\bibinfo{author}{Xu, Z.}, \bibinfo{author}{Zhu, X.}, \bibinfo{author}{Zhao, G.-D.}, \bibinfo{author}{Zhang, D.~W.} \& \bibinfo{author}{Yu, S.}
\newblock \bibinfo{title}{Oxygen vacancies stabilized 180$^\circ$ charged domain walls in ferroelectric hafnium oxide}.
\newblock \emph{\bibinfo{journal}{Applied Physics Letters}} \textbf{\bibinfo{volume}{124}} (\bibinfo{year}{2024}).

\bibitem{noor2024nearly}
\bibinfo{author}{Noor, M.} \emph{et~al.}
\newblock \bibinfo{title}{Nearly barrierless polarization switching mechanisms in zro2 having perpendicular in-plane domain walls}.
\newblock \emph{\bibinfo{journal}{ACS Applied Materials \& Interfaces}} \textbf{\bibinfo{volume}{16}}, \bibinfo{pages}{62282--62291} (\bibinfo{year}{2024}).

\bibitem{shin2007nucleation}
\bibinfo{author}{Shin, Y.-H.}, \bibinfo{author}{Grinberg, I.}, \bibinfo{author}{Chen, I.-W.} \& \bibinfo{author}{Rappe, A.~M.}
\newblock \bibinfo{title}{Nucleation and growth mechanism of ferroelectric domain-wall motion}.
\newblock \emph{\bibinfo{journal}{Nature}} \textbf{\bibinfo{volume}{449}}, \bibinfo{pages}{881--884} (\bibinfo{year}{2007}).

\bibitem{garrity2014hyperferroelectrics}
\bibinfo{author}{Garrity, K.~F.}, \bibinfo{author}{Rabe, K.~M.} \& \bibinfo{author}{Vanderbilt, D.}
\newblock \bibinfo{title}{Hyperferroelectrics: proper ferroelectrics with persistent polarization}.
\newblock \emph{\bibinfo{journal}{Physical review letters}} \textbf{\bibinfo{volume}{112}}, \bibinfo{pages}{127601} (\bibinfo{year}{2014}).

\bibitem{catalan2012domain}
\bibinfo{author}{Catalan, G.}, \bibinfo{author}{Seidel, J.}, \bibinfo{author}{Ramesh, R.} \& \bibinfo{author}{Scott, J.~F.}
\newblock \bibinfo{title}{Domain wall nanoelectronics}.
\newblock \emph{\bibinfo{journal}{Reviews of Modern Physics}} \textbf{\bibinfo{volume}{84}}, \bibinfo{pages}{119--156} (\bibinfo{year}{2012}).

\bibitem{meier2012anisotropic}
\bibinfo{author}{Meier, D.} \emph{et~al.}
\newblock \bibinfo{title}{Anisotropic conductance at improper ferroelectric domain walls}.
\newblock \emph{\bibinfo{journal}{Nature materials}} \textbf{\bibinfo{volume}{11}}, \bibinfo{pages}{284--288} (\bibinfo{year}{2012}).

\bibitem{zhang2013direct}
\bibinfo{author}{Zhang, Q.} \emph{et~al.}
\newblock \bibinfo{title}{Direct observation of multiferroic vortex domains in ymno3}.
\newblock \emph{\bibinfo{journal}{Scientific reports}} \textbf{\bibinfo{volume}{3}}, \bibinfo{pages}{2741} (\bibinfo{year}{2013}).

\bibitem{yang2025enhanced}
\bibinfo{author}{Yang, B.} \emph{et~al.}
\newblock \bibinfo{title}{Enhanced energy storage in antiferroelectrics via antipolar frustration}.
\newblock \emph{\bibinfo{journal}{Nature}} \textbf{\bibinfo{volume}{637}}, \bibinfo{pages}{1104--1110} (\bibinfo{year}{2025}).

\bibitem{zhong2026observation}
\bibinfo{author}{Zhong, H.} \emph{et~al.}
\newblock \bibinfo{title}{Observation of one-dimensional, charged domain walls in ferroelectric zro2}.
\newblock \emph{\bibinfo{journal}{Science}} \textbf{\bibinfo{volume}{391}}, \bibinfo{pages}{407--411} (\bibinfo{year}{2026}).

\bibitem{mukundan2019structural}
\bibinfo{author}{Mukundan, V.} \emph{et~al.}
\newblock \bibinfo{title}{Structural correlation of ferroelectric behavior in mixed hafnia-zirconia high-k dielectrics for feram and ncfet applications}.
\newblock \emph{\bibinfo{journal}{MRS Advances}} \textbf{\bibinfo{volume}{4}}, \bibinfo{pages}{545--551} (\bibinfo{year}{2019}).

\bibitem{hachtel2018sub}
\bibinfo{author}{Hachtel, J.~A.}, \bibinfo{author}{Idrobo, J.~C.} \& \bibinfo{author}{Chi, M.}
\newblock \bibinfo{title}{Sub-{\aa}ngstrom electric field measurements on a universal detector in a scanning transmission electron microscope}.
\newblock \emph{\bibinfo{journal}{Advanced structural and chemical imaging}} \textbf{\bibinfo{volume}{4}}, \bibinfo{pages}{10} (\bibinfo{year}{2018}).

\bibitem{nord2017atomap}
\bibinfo{author}{Nord, M.}, \bibinfo{author}{Vullum, P.~E.}, \bibinfo{author}{MacLaren, I.}, \bibinfo{author}{Tybell, T.} \& \bibinfo{author}{Holmestad, R.}
\newblock \bibinfo{title}{Atomap: a new software tool for the automated analysis of atomic resolution images using two-dimensional gaussian fitting}.
\newblock \emph{\bibinfo{journal}{Advanced structural and chemical imaging}} \textbf{\bibinfo{volume}{3}}, \bibinfo{pages}{1--12} (\bibinfo{year}{2017}).

\bibitem{kresse1996efficient}
\bibinfo{author}{Kresse, G.} \& \bibinfo{author}{Furthm{\"u}ller, J.}
\newblock \bibinfo{title}{Efficient iterative schemes for ab initio total-energy calculations using a plane-wave basis set}.
\newblock \emph{\bibinfo{journal}{Physical review B}} \textbf{\bibinfo{volume}{54}}, \bibinfo{pages}{11169} (\bibinfo{year}{1996}).

\bibitem{kresse1996efficiency}
\bibinfo{author}{Kresse, G.} \& \bibinfo{author}{Furthm{\"u}ller, J.}
\newblock \bibinfo{title}{Efficiency of ab-initio total energy calculations for metals and semiconductors using a plane-wave basis set}.
\newblock \emph{\bibinfo{journal}{Computational materials science}} \textbf{\bibinfo{volume}{6}}, \bibinfo{pages}{15--50} (\bibinfo{year}{1996}).

\bibitem{kresse1993ab}
\bibinfo{author}{Kresse, G.} \& \bibinfo{author}{Hafner, J.}
\newblock \bibinfo{title}{Ab initio molecular dynamics for liquid metals}.
\newblock \emph{\bibinfo{journal}{Physical review B}} \textbf{\bibinfo{volume}{47}}, \bibinfo{pages}{558} (\bibinfo{year}{1993}).

\bibitem{blochl1994projector}
\bibinfo{author}{Bl{\"o}chl, P.~E.}
\newblock \bibinfo{title}{Projector augmented-wave method}.
\newblock \emph{\bibinfo{journal}{Physical review B}} \textbf{\bibinfo{volume}{50}}, \bibinfo{pages}{17953} (\bibinfo{year}{1994}).

\bibitem{perdew2008restoring}
\bibinfo{author}{Perdew, J.~P.} \emph{et~al.}
\newblock \bibinfo{title}{Restoring the density-gradient expansion for exchange in solids and surfaces}.
\newblock \emph{\bibinfo{journal}{Physical review letters}} \textbf{\bibinfo{volume}{100}}, \bibinfo{pages}{136406} (\bibinfo{year}{2008}).

\bibitem{monkhorst1976special}
\bibinfo{author}{Monkhorst, H.~J.} \& \bibinfo{author}{Pack, J.~D.}
\newblock \bibinfo{title}{Special points for brillouin-zone integrations}.
\newblock \emph{\bibinfo{journal}{Physical review B}} \textbf{\bibinfo{volume}{13}}, \bibinfo{pages}{5188} (\bibinfo{year}{1976}).

\bibitem{henkelman2000improved}
\bibinfo{author}{Henkelman, G.} \& \bibinfo{author}{J{\'o}nsson, H.}
\newblock \bibinfo{title}{Improved tangent estimate in the nudged elastic band method for finding minimum energy paths and saddle points}.
\newblock \emph{\bibinfo{journal}{The Journal of chemical physics}} \textbf{\bibinfo{volume}{113}}, \bibinfo{pages}{9978--9985} (\bibinfo{year}{2000}).

\bibitem{gonze1994interatomic}
\bibinfo{author}{Gonze, X.}, \bibinfo{author}{Charlier, J.-C.}, \bibinfo{author}{Allan, D.} \& \bibinfo{author}{Teter, M.}
\newblock \bibinfo{title}{Interatomic force constants from first principles: The case of $\alpha$-quartz}.
\newblock \emph{\bibinfo{journal}{Physical Review B}} \textbf{\bibinfo{volume}{50}}, \bibinfo{pages}{13035} (\bibinfo{year}{1994}).

\bibitem{togo2015first}
\bibinfo{author}{Togo, A.} \& \bibinfo{author}{Tanaka, I.}
\newblock \bibinfo{title}{First principles phonon calculations in materials science}.
\newblock \emph{\bibinfo{journal}{Scripta Materialia}} \textbf{\bibinfo{volume}{108}}, \bibinfo{pages}{1--5} (\bibinfo{year}{2015}).

\end{thebibliography}

\newpage

%\documentclass{WileyMSP-template}
%\usepackage{xcolor}
%\definecolor{dark_green}{RGB}{0,120,0}
%\usepackage{amsmath}
%\usepackage{textcomp}
%\usepackage{comment}
%\usepackage{siunitx}
%\begin{document}
%\begin{comment}
%\pagestyle{fancy}
%\rhead{\includegraphics[width=2.5cm]{vch-logo.png}}

%\title{Sub-nm\textsuperscript{2} ferroelectric domains via charged 180° walls in ZrO\textsubscript{2}}

%\title{Supporting Information}

\section{Supporting Information}
\title{Supporting Information}
\makeatletter
\setcounter{figure}{0}
\renewcommand{\thefigure}{S\arabic{figure}}
\renewcommand{\figurename}{Figure}
\renewcommand{\fnum@figure}{\figurename~\thefigure}
%\makeatother
%\renewcommand{\thefigure}{E\arabic{figure}}%
%\clearpage

%\justifying
%\title{\vspace{-1.0em}\Large Supporting Information\\[0.4em]

%\subsection{Figure S1.}
The overview of the Al/ TiN/ ZrO\textsubscript{2}/SiO\textsubscript{2}/Si device structure along with the cross-sectional TEM image is shown in Figure S1a,b. The PV loop obtained from the out-of-plane metal-insulator-metal device shows the anti-clockwise hysteresis.   

\begin{figure}[ht]
\centering
\includegraphics[width=6.5in]{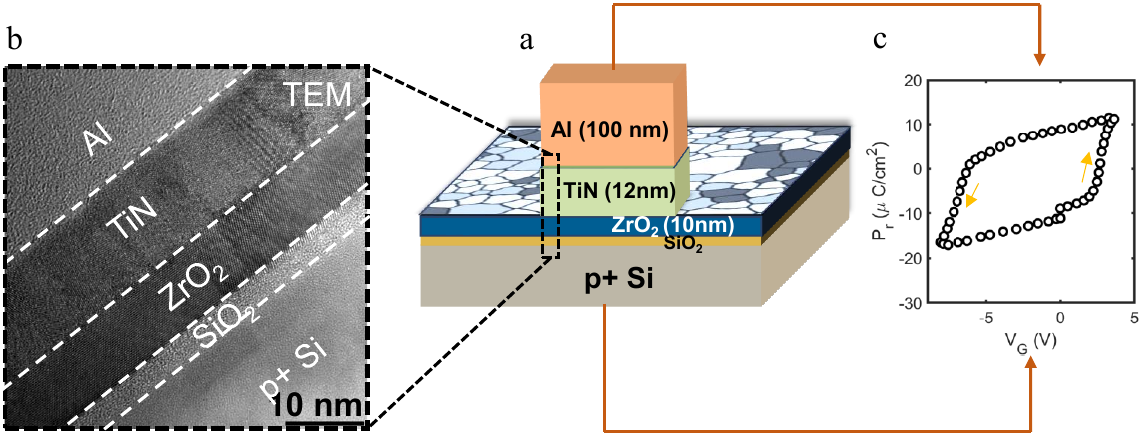}\vspace{-.0in}
\caption{\textbf{Electrical characterization of ZrO\textsubscript{2} MFS sample.} \textbf{(a)} Overview of the Al/TiN/ZrO\textsubscript{2}/SiO\textsubscript{2}/p+ Si thin film heterostructure. \textbf{(b)} HRTEM image of the thin film heterostructure. \textbf{c} P-V loop obtained from the heterostructure confirming the ferroelectricity of the device.}
\label{fig_S1}
\end{figure}

\newpage
%\subsection{Figure S2.}
After electrical characterization and removing top electrodes via wet-
etch, Grazing incidence x-ray diffraction (GIXRD) curve was obtained for devices with 10 nm ferroelectric ZrO\textsubscript{2} layer, which is shown in Figure S2. It confirms strong crystallinity and the presence of dominant orthorhombic ferroelectric phase in the film. 

\begin{figure}[ht]
\centering
\includegraphics[width=5.5in]{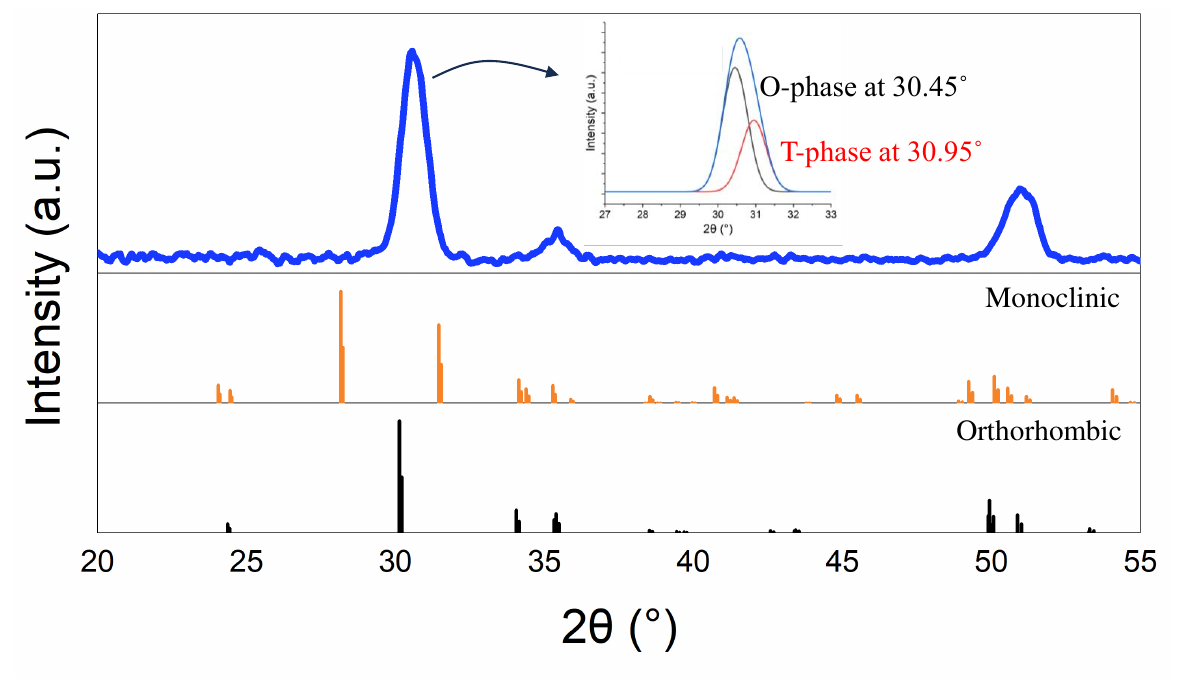}\vspace{-.0in}
\caption{\textbf{GIXRD scans for 10 nm thick zirconia-based thin film capacitor conforming the orthorhombic phase of the film.} Inset shows the de-convolution of dominant peak which has O-phase ($Pca2_1$ phase) as majority.}
\label{fig_S1}
\end{figure}

\newpage
%\subsection{Figure S3.}
Planar (i.e., plan-view or top-down) TEM samples are prepared according to the process flow shown in Figure S3a. After etching the top TiN electrode, FE
ZrO\textsubscript{2}/SiO\textsubscript{2}/Si thin film heterostructures from a bulk wafer sample are disc cut into 3 mm diameter discs using a femto-laser cutter. The sample disc is then adhered to a steel polishing stub with the silicon side facing up. The Si is manually polished to a thickness of approximately 60-80 $\mu$m using a Gatan disc grinder and a series of SiC polishing paper and diamond polishing pastes. Si at the middle region is further thinned down to $\sim 10\,\mu m$ using dimple grinder. Finally, the sample is ion milled using a Gatan Model 691 Precision Ion Polishing System (PIPS). Top, side, and bottom view schematics of the final plan-view sample are also depicted in Figure S3b.

\begin{figure}[ht]
\centering
\includegraphics[width=6.5in]{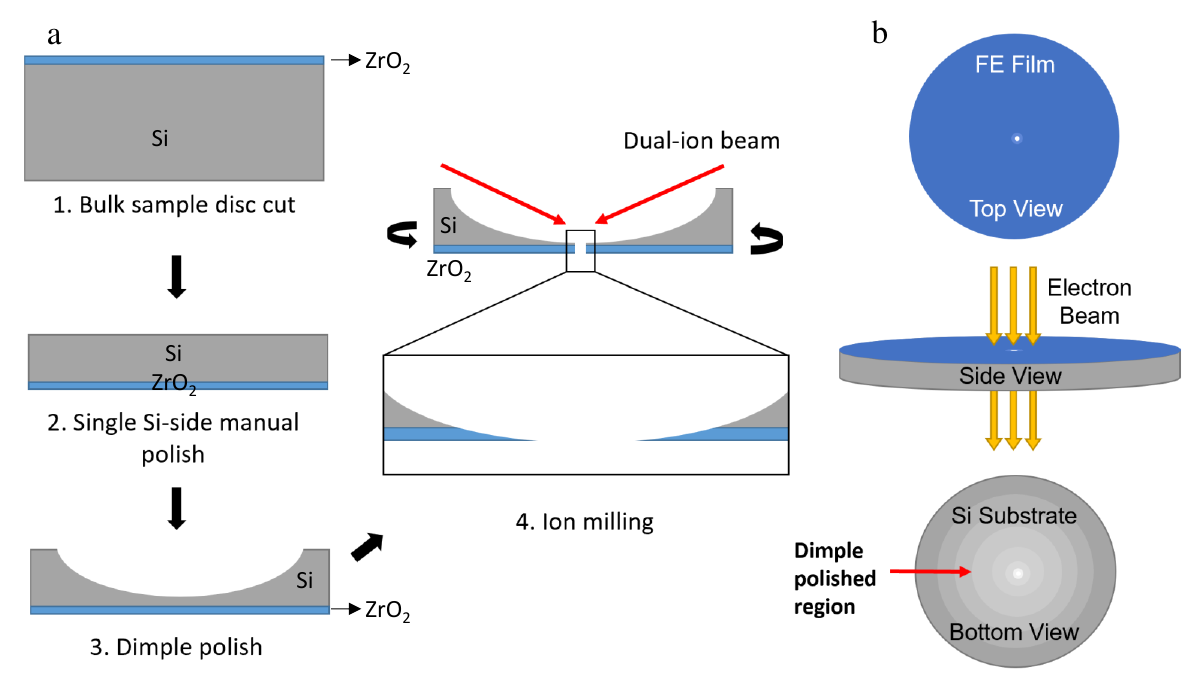}\vspace{-.0in}
\caption{\textbf{Plan-view ion milling TEM sample preparation process flow.} \textbf{(a)} (1) The bulk sample is cut into discs of 3 mm diameter, and (2) the Si substrate is thinned down to 80-100 \(\mu\)m using manual polisher. (3) Dimple polisher is used thin down the sample from Si side to 10 \(\mu\)m, (4) followed by Ar ion milling to remove Si and expose the ZrO\textsubscript{2} film in the center of the discs. \textbf{(b)} Top, side, and bottom views of the sample after ion milling.}
\label{fig_S2}
\end{figure}

%\subsection{Extended Data Figure S4.}
%The displacement of O\textsubscript{II} along [00-1] direction from the centro-symmetric position across each TT and HH domains of Figure S 2d is shown in extended data Figure S 4 along with error bars. The ABF image of individual HH and TT domains are shown in extended data Figure S 5 and 7, and O\textsubscript{II} displacements along those domains which are used to construct this plot are shown in extended data Figure S 6 and 8 respectively. The zero value on the x-axis represents the domain wall region. At the TT boundary, O has almost zero displacement at each polar row. On the other hand, the domain boundary of HH domains doesn't show any O atom, that is, the O which is aligning with Zr atoms in the HH boundary region is not showing up in \textbf{a}. This is because those atoms cannot be mapped with four neighboring Zr atoms forming the cage. It is to be noted that, the displacement of O is maximum at the adjacent unit cells of the HH domain wall. 

\newpage
%\subsection{Figure S4.}
An overview of the imaging setup is shown in Figure S4a, through which nano to atomic-scale images reveal the in-plane grain morphology and precise locations of Zr and O atoms in the ZrO\textsubscript{2} thin film. It shows a low magnification STEM image of the film capturing multiple discernible grains, each extending from a few to a hundred nanometers with varying phases and orientations. Figure S4b shows a HAADF image revealing the positions of Zr atoms while the dipole directions can be obtained from the ABF image (Figure S4c), which resolves both Zr and O atomic columns.

\begin{figure}[ht]
\centering
\includegraphics[width=4.5in]{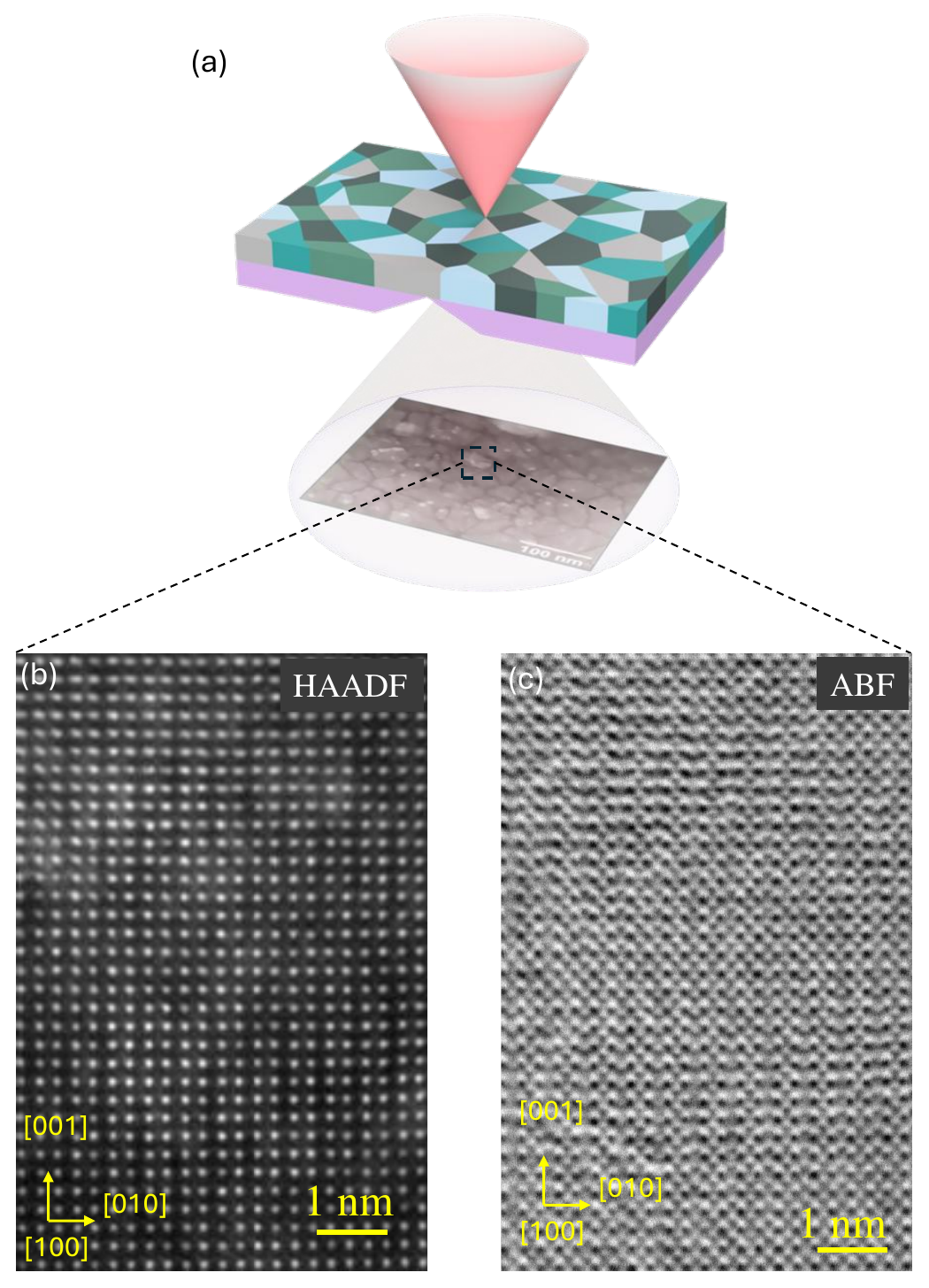}\vspace{-.0in}
\caption{\textbf{STEM imaging of ZrO\textsubscript{2} plan-view sample.} \textbf{(a)} An overview of the imaging setup for imaging in-plane domains of the 10 nm ZrO\textsubscript{2} plan-view sample. \textbf{(b)} HAADF and \textbf{(c)} ABF images captured simultaneously to image both Zr and O atoms from one of the grains of ZrO\textsubscript{2} having [100] zone axis.}
\label{fig_S2}
\end{figure}

\newpage
%\subsection{Figure S5.}
The simulated ABF images of O-phase ($Pca2_1$) with opposite polarization, as shown in Figure S5a,b, were used as templates to detect local structural correlations of polar domains with opposite polarization across the experimental ABF image in Figure S5c. The spatial variation of structural similarities with polar domains, assessed through template-matching, are depicted as color maps in Figure S5d. Here, the red color indicates a high similarity of the polar structure with polarization pointing upward (+1), as shown in Figure S5a, while the blue color corresponds to polarization pointing downward (-1), as shown in Figure S5b. Based on the template-matching analysis, local structures exhibiting maximum correlation with the simulated ABF images were extracted from the ABF image of Figure S5c. As shown in the bottom panels of Figure S5a,b, the extracted local domains from the experimental ABF images show excellent agreement with the simulated ABF templates shown in the top panels.

\begin{figure}[ht]
\centering
\includegraphics[width= 6.5in]{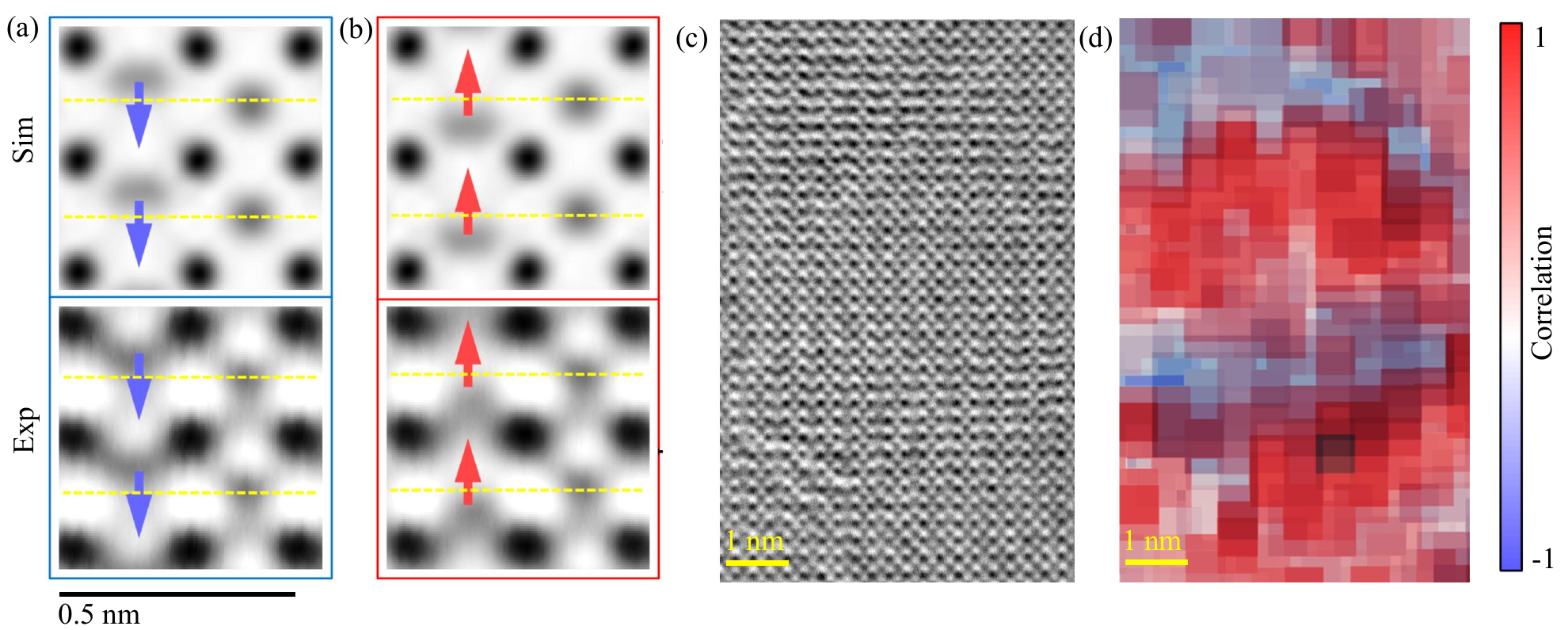}\vspace{-.0in}
\caption{\textbf{Local structural correlation analysis for polar domains with opposite polarization.} \textbf{(a,b)} Simulated and experimental ABF images for polarization pointing downward in panel (a) and pointing upward in panel (b). \textbf{(c)} Experimental ABF image. \textbf{(d)} Template-matching using fast normalized cross-correlation showing the local structural correlation for different polar states in panels (a) and (b). Color bars quantify the cross-correlation with the template images having different polar states. }
\label{fig_S2}
\end{figure}

\newpage
%\subsection{Figure S6.}

\begin{figure}[ht]
\centering
\includegraphics[width=6.5in]{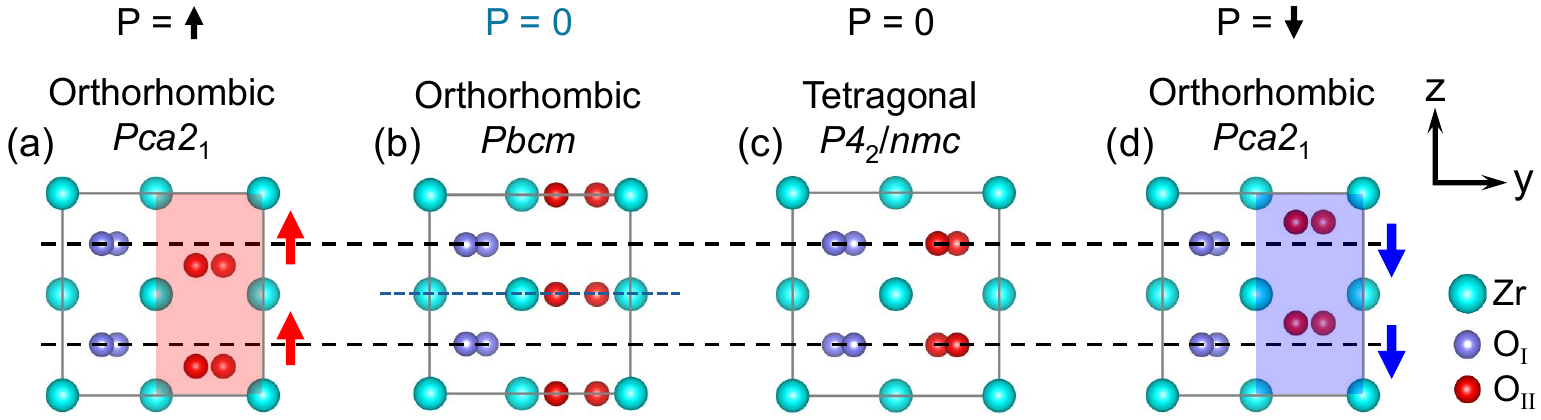}\vspace{-.0in}
\caption{\textbf{The unit cell structures of three different phases of ZrO$_2$} \textbf{(a)} polar orthorhombic $Pca2_1$ (up-polarized), two competing switching intermediate states of \textbf{(b)} nonpolar orthorhombic $Pbcm$ and \textbf{(c)} nonpolar tetragonal $P4_2/nmc$, and \textbf{(d)} polar orthorhombic $Pca2_1$ (down-polarized). Note that, considering the high-symmetry cubic phase as the reference, in a transition from the up-polarized $Pca2_1$ structure shown in panel (a) to the down-polarized $Pca2_1$ structure shown in panel (d), the phonon modes of $\Gamma_{15}^{z}$, $X_{2}^{\prime}$, $Y_{5}^{z}$, and $Z_{5}^{x}$ reverse their signs.}
\label{fig_S1}
\end{figure}

\newpage
%\subsection{Figure S7.}
Figure S7 shows ABF images of two different (100) regions in ZrO\textsubscript{2} thin film which contains HH domains. ABF images of Figure 1 are taken from these grains. Every polar layer which has HH boundary are numbered in Figure S7a,b. Each polar layer from the images is shown separately in the bottom part of the figure. The O\textsubscript{II} atoms getting lined up with Zr atoms are evident in each boundary region. 

\begin{figure}[ht]
\centering
\includegraphics[width=5.5 in, height=7in]{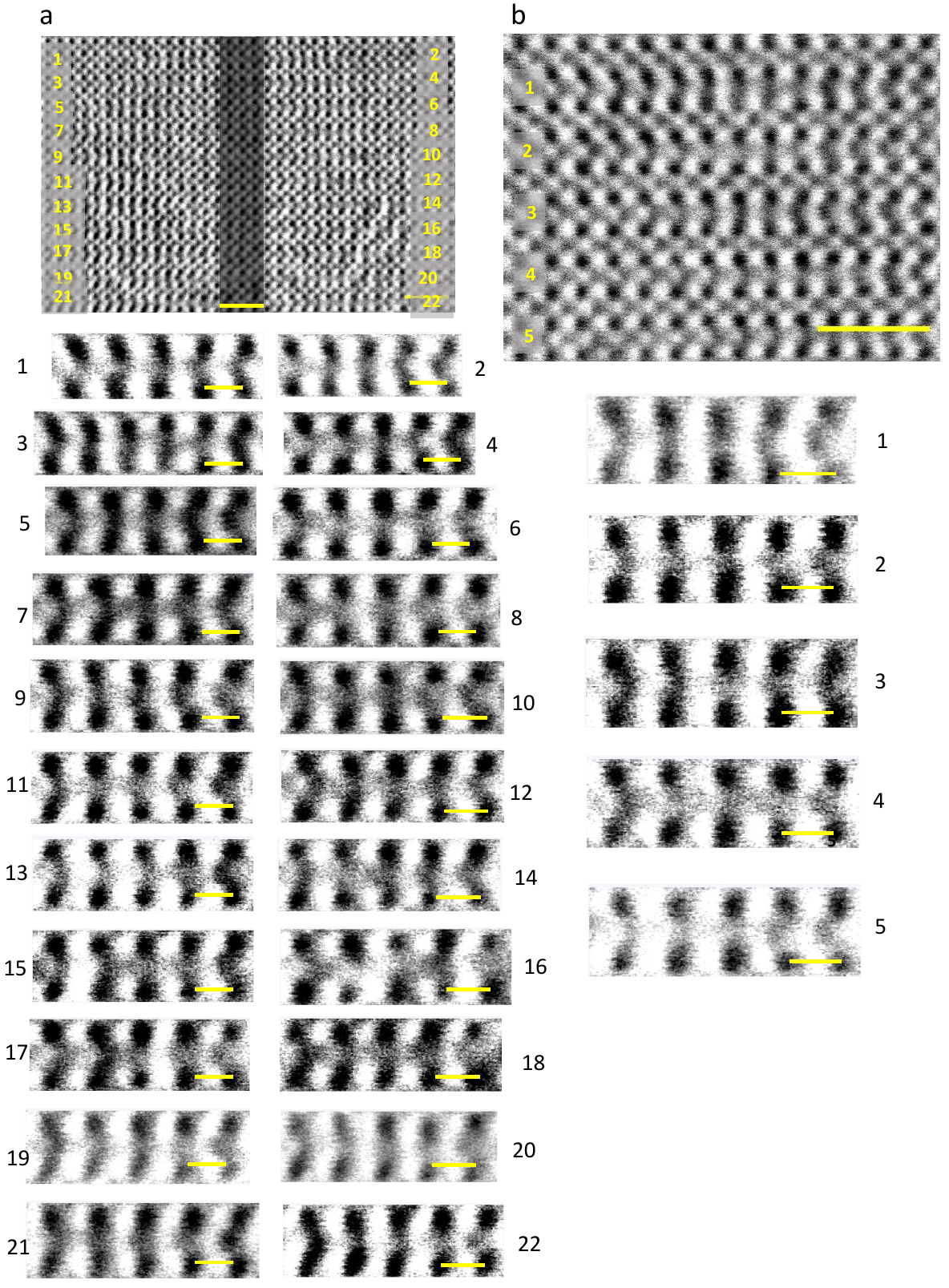}\vspace{-.0in}
\caption{\textbf{ABF images of [100] zone axis of 10nm ZrO\textsubscript{2}.}  Domain walls of all the individual HH domain columns of the image are shown separately. The scale bars of \textbf{(a)} and \textbf{(b)} are 1nm and those of individual columns are 0.3 nm.}
\label{fig_S3}
\end{figure}

\newpage
%\subsection{Figure S8.}
A grain of ZrO\textsubscript{2} film in [100] Pca2\textsubscript{1} orientation is shown in Figure S8 containing polarization directions. Blue and red arrows represent right pointing and left pointing domains, respectively. The lengths of the arrows depend on the amount of O displacement from the centro-symmetric position. All the atomic positions were extracted using Atomap library in Python, and O\textsubscript{II} displacements were calculated accordingly. These O\textsubscript{II} displacements are plotted separately for each polar layer of HH domains as numbered in the figure. In each graph, the blue dots are coming from the blue polarization vectors and red dots are coming from red polarization vectors. The domain boundary of HH domains doesn't show any O atom, that is, the O\textsubscript{II} atoms which are aligning with Zr atoms in the HH boundary region are not showing up in the O\textsubscript{II} displacement plots. This is because those atoms cannot be mapped with four neighboring Zr atoms forming the cage. It is to be noted that, the displacement of O is maximum at the adjacent unit cells of the HH domain wall. 
\newpage
\begin{figure}[ht]
\centering
\includegraphics[width=5.5in, height= 6.5 in]{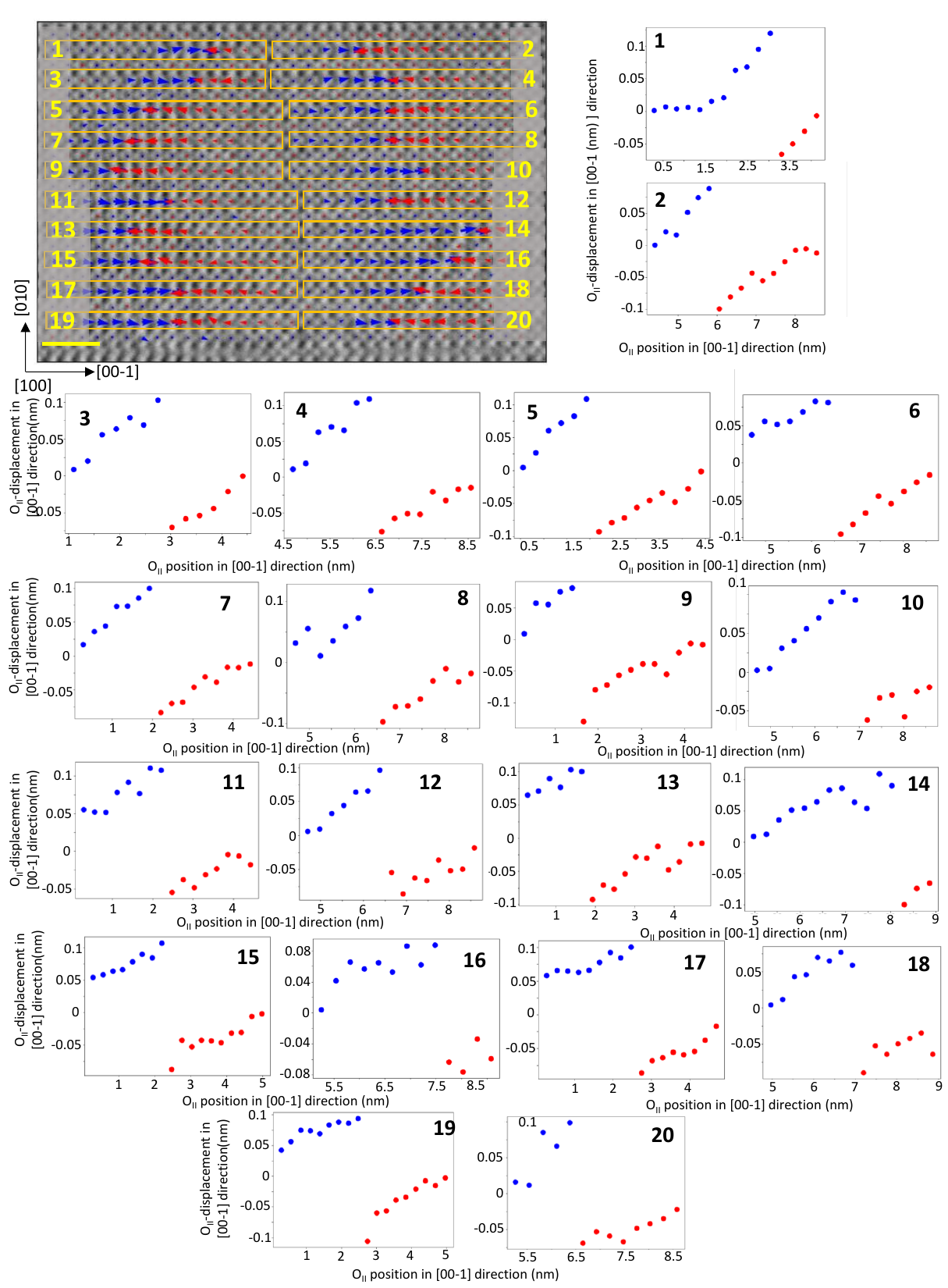}\vspace{-.0in}
\caption{\textbf{O\textsubscript{II} displacement along the HH domains.} The displacement of O\textsubscript{II} from its centro-symmetric positions within all HH domains (highlighted by yellow boxes) along the [00$\bar{1}$] direction is shown separately, with each domain numbered. In the graphs, the O\textsubscript{II} displacement within the right-pointing domains (blue) is considered positive. Scale bar = 1 nm.}
\vspace{-.0in}
\label{Fig_S4}
\end{figure}

%\subsection{Extended Data Figure S 5.}
%Extended data Figure S5 shows ABF images of two different (100) regions in ZrO\textsubscript{2} thin film which contains TT domains. First grain is the same region shown in the extended data Figure S3a, and second grain is a portion of the grain shown in of extended data Figure S4. Every polar layer which has TT boundary are numbered in the grains. Each polar layer from the images is shown separately in the bottom part of the figure. O residing at the centrosymmetric position of four Zr atoms are seen at each boundary.

\newpage
%\subsection{Figure S9.}
Figure S9 shows ABF images of [100] region in ZrO\textsubscript{2} thin film which contains TT domains. The grain is the same region shown in the Figure S7a. Every polar layer which has TT domains are numbered in the grain. Each polar layer from the image is shown separately at the bottom part of the figure. O\textsubscript{II} residing at the centrosymmetric position of four Zr atoms (Zr cage) are seen at each boundary.

\begin{figure}[ht]
\centering
\includegraphics[width=2.5 in, height= 6.5 in]{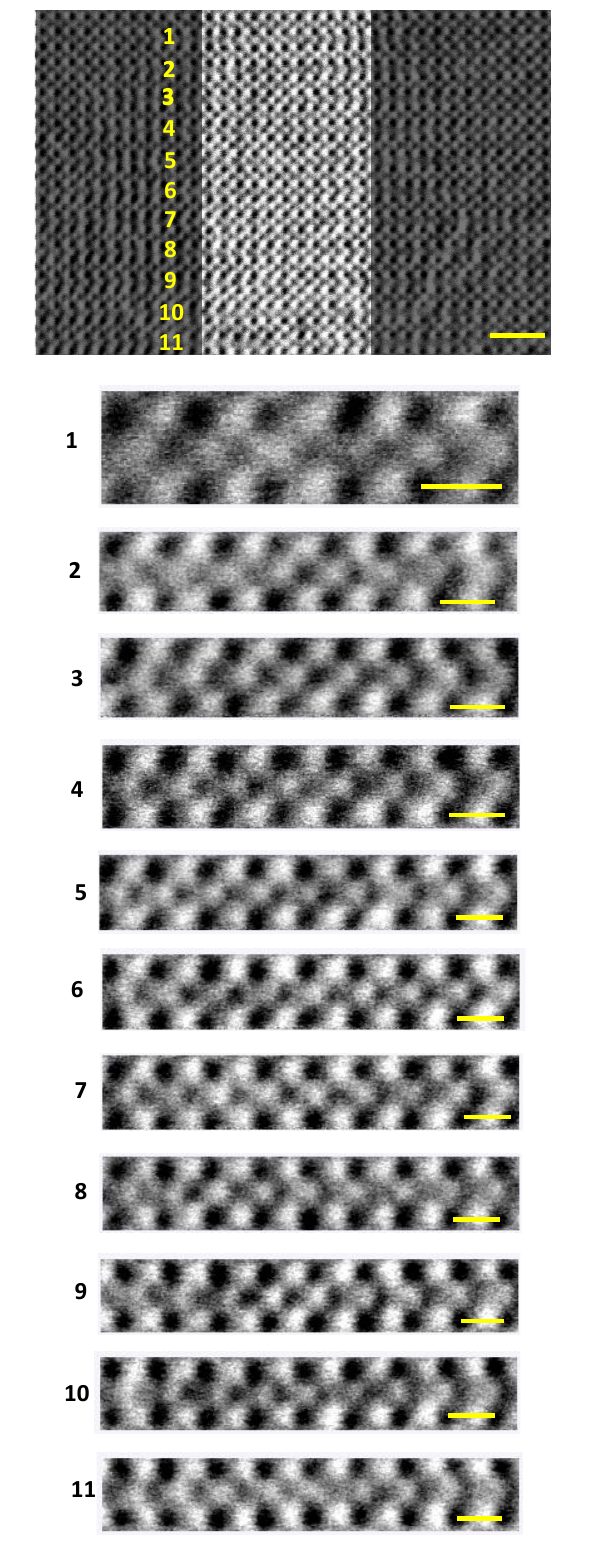}
\caption{\textbf{ABF images along [100] zone axis of 10 nm ZrO\textsubscript{2}.} Domain walls of all the individual TT domain columns of the image are shown separately. The scale bars of the full grains are 1 nm and those of individual columns are 0.3 nm.}
\label{fig_S5}
\end{figure}

\newpage
%\subsection{Figure S10.}
Figure S10 shows the same grain as in Figure S8. Each graph represents the magnitude of O\textsubscript{II} displacement from centrosymmetric position along the TT domains. It is seen that, at the TT boundary, O has almost zero displacement at each polar row.

\begin{figure}[ht]
\centering
\includegraphics[width=5in, height= 6.5 in]{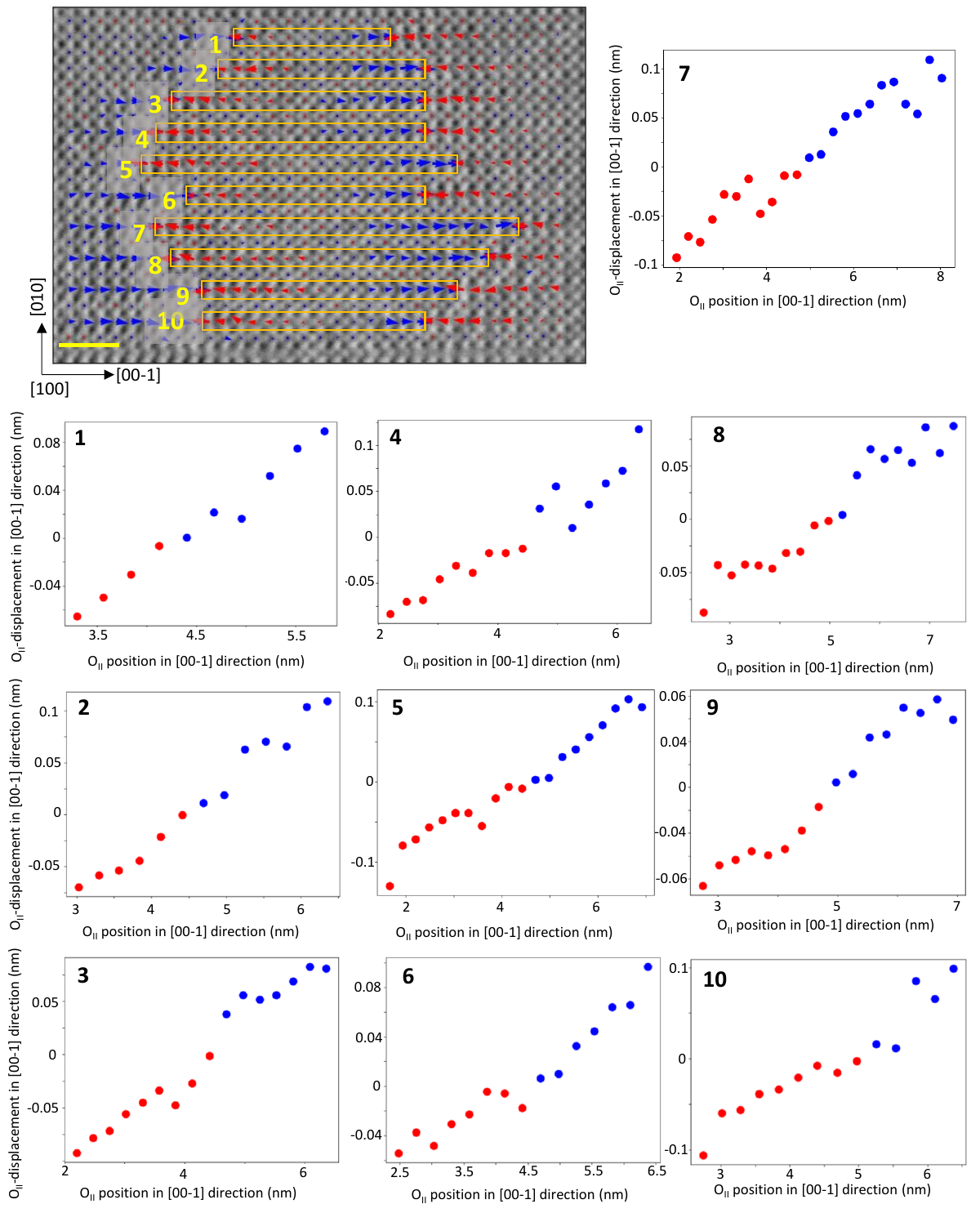}
\caption{\textbf{O\textsubscript{II} displacement across the TT domains.} The displacement of O\textsubscript{II} from its centro-symmetric positions within all TT domains (highlighted by yellow boxes) along the [00$\bar{1}$] direction is shown separately, with each domain numbered. In the graphs, the O\textsubscript{II} displacement within the right-pointing domains (blue) is considered positive. Scale bar = 1 nm.}
\label{fig_S6}
\end{figure}

\newpage
%\subsection{Figure S11.}
By quantifying the polar displacements from the ABF images, as shown in Figures S11a,b, we plotted the off-center displacements of the oxygen atomic columns as a function of their horizontal position in the image within the same polar layers, as shown in Figure S11c. The experimental measurements show good statistical agreement with the DFT-simulated results further validating the structure of the domain walls.

\begin{figure}[ht]
\centering
\includegraphics[width=6in]{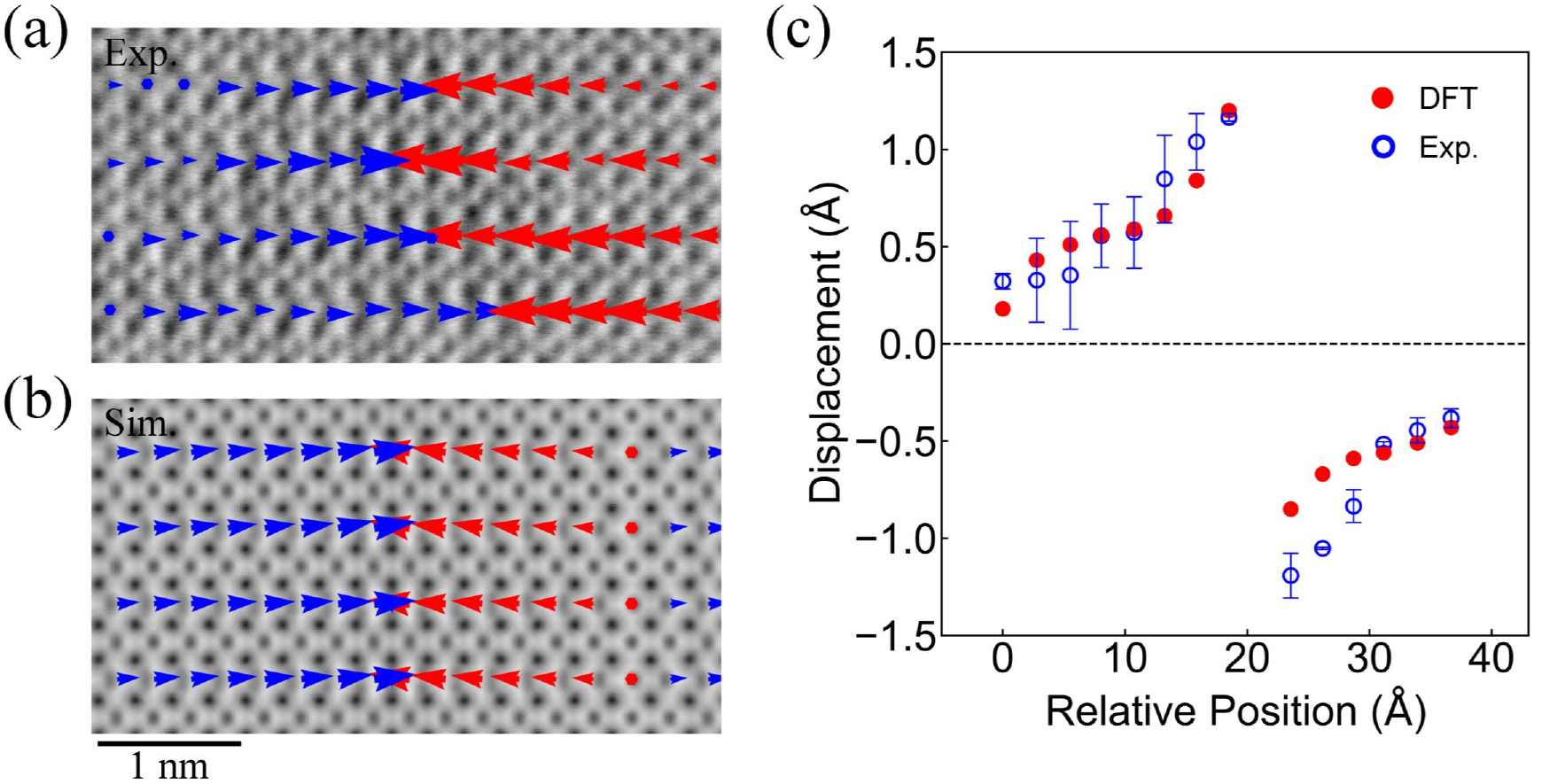}
\caption{\textbf{Polar displacement of oxygen atomic columns across the charged 180° domain walls.} Polar displacement measurements from experimental \textbf{(a)} and simulated \textbf{(b)} ABF images. \textbf{(c)} Statistical distribution of oxygen displacement magnitudes as a function of oxygen position within a polar layer. Error bars represent the standard deviation of displacements across all polar layers measured in (a).}
\label{fig_S6}
\end{figure}

\newpage
\begin{figure}[ht]
\centering
\includegraphics[width=6.4in]{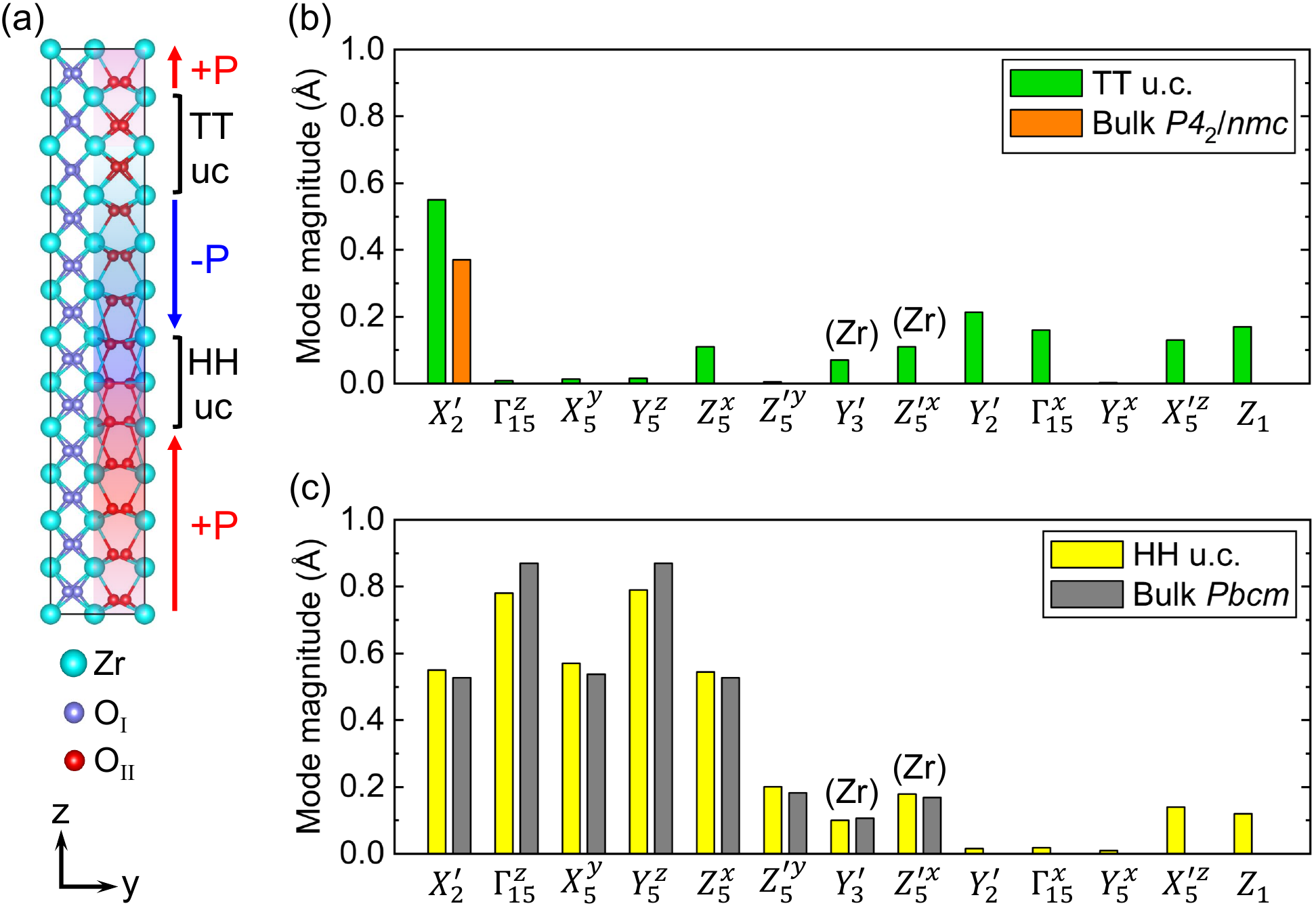}
\caption{\textbf{Phase identification at the HH and TT charged domain walls in ZrO\textsubscript{2}.} \textbf{(a)} Fully relaxed atomic structure of a closely spaced HH–TT configuration. \textbf{(b)} and \textbf{(c)} Absolute amplitudes of the phonon modes condensed at the TT and HH regions (twelve-atom unit cells), compared with those of the bulk $P4_2/nmc$ and $Pbcm$ structures, respectively. For the HH-wall unit cell, two oxygen atoms were removed when performing the phonon decomposition. For all phonon-mode decompositions, the high-symmetry cubic phase was used as the reference. The result verifies that the TT and HH regions feature distorted $P4_2/nmc$ and $Pbcm$ phases, respectively.}
\label{fig_S12}
\end{figure}

\newpage
\begin{figure}[ht]
\centering
\includegraphics[width=5in]{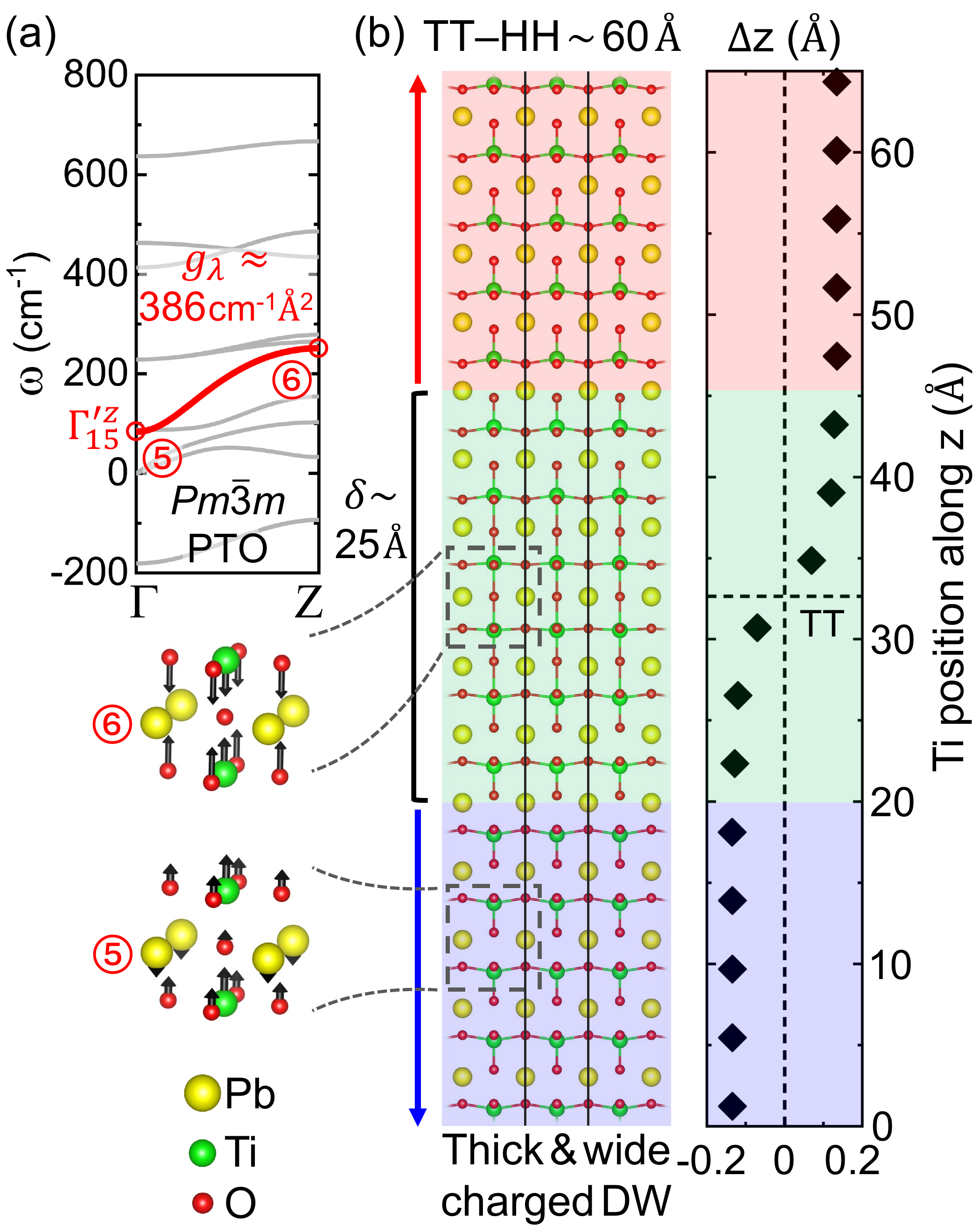}
\caption{\textbf{Largely dispersed LO polar band in PTO and its thick and wide charged 180\(^{\circ}\) domain walls.} \textbf{(a)} Phonon band dispersion for PTO cubic structure. The eigenvectors corresponding to the LO polar mode at $\Gamma$-point and its respective band-connected mode at $Z$-point are visualized (\textcircled{5} and \textcircled{6}) by black arrows. \textbf{(b)} TT-wall region of the DFT-optimized PTO HH-TT configuration and the related off-center polar displacements of Ti atoms. Note that within the domain (bulk-like) regions, the Ti off-center polar displacements are approximately 0.135 \AA, whereas in the ideal bulk structure they are 0.158 \AA, implying an approximately 15\% suppression of polarization in the domains of PTO CDW system. The full HH-TT structure is shown in the next figure}
\label{fig_S13}
\end{figure}

\newpage
\begin{figure}[ht]
\centering
\includegraphics[width=7.4in]{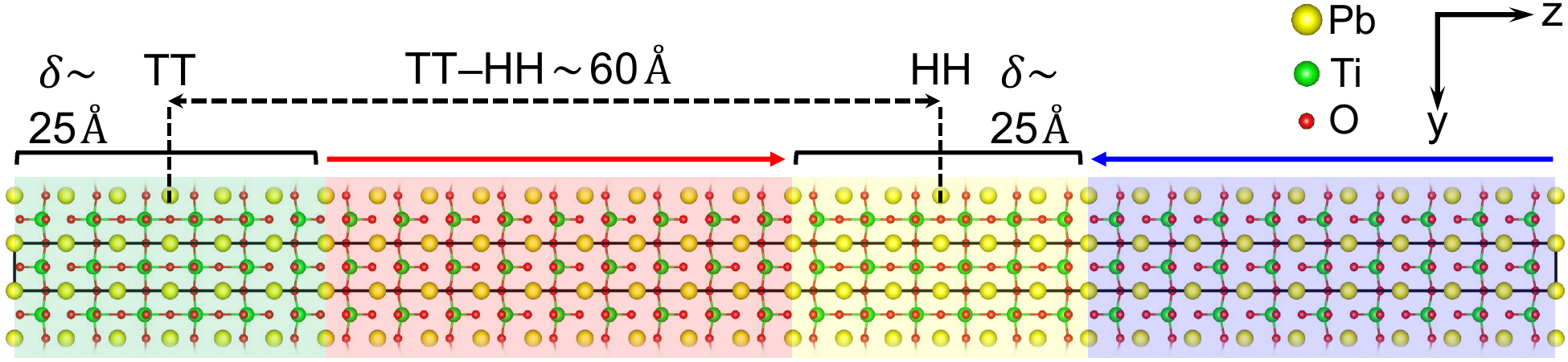}
\caption{\textbf{The relaxed PTO HH–TT configuration with a total length of 12.5 nm. It features charged 180° domain walls with an approximate thickness of 2.5 nm, along with bulk-like domain regions extending over nearly 3.75 nm.}}
\label{fig_S14}
\end{figure}

\newpage
\begin{figure}[ht]
\centering
\includegraphics[width=4.5in]{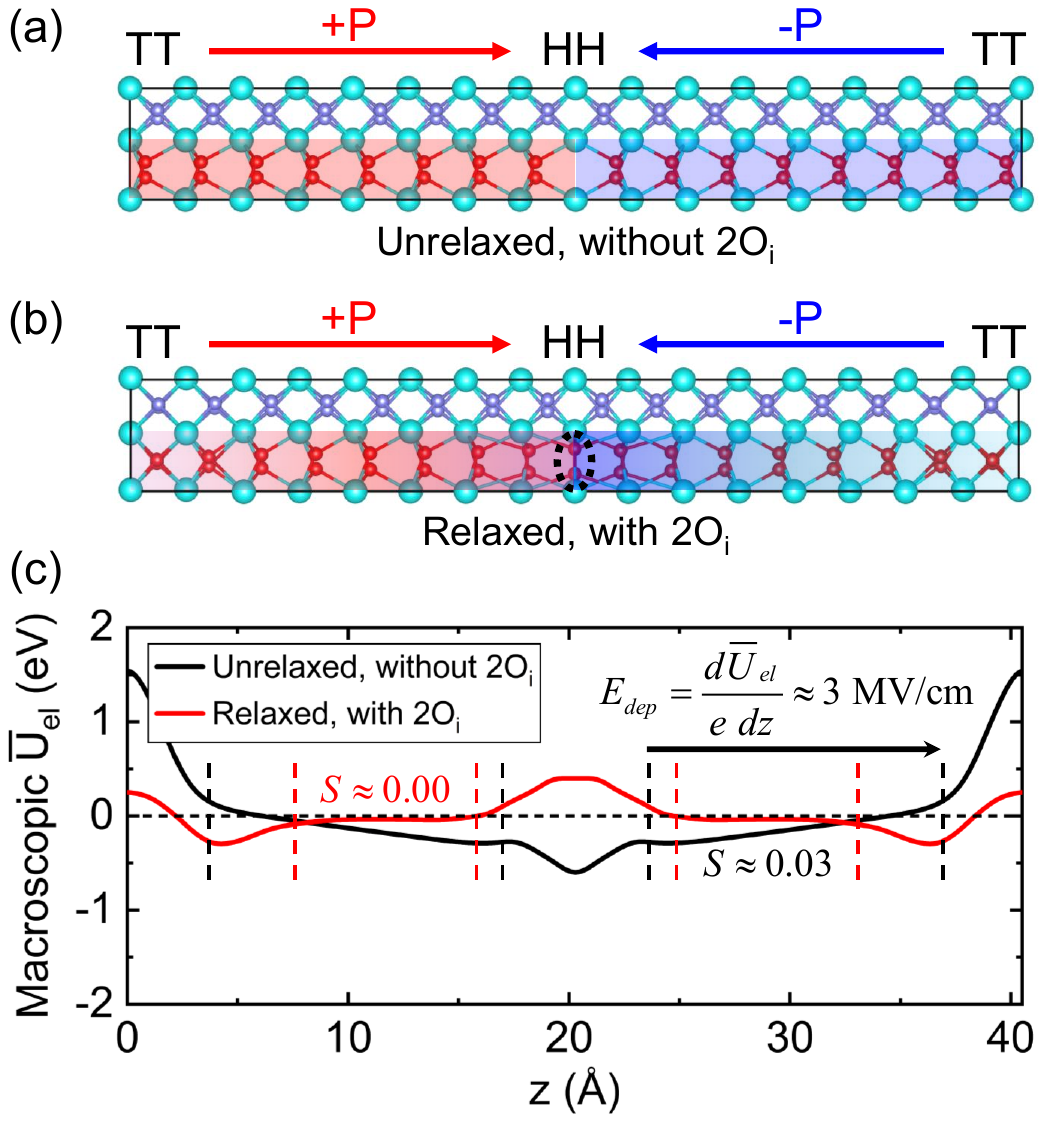}
\caption{\textbf{Depolarization field ($E_{dep}$) suppression by two interstitial oxygen atoms, naturally located at the HH wall.} \textbf{(a)} Unrelaxed domain wall structure without O\textsubscript{i}, \textbf{(b)} relaxed domain wall structure containing two O\textsubscript{i}, and \textbf{(c)} their macroscopic average electrostatic potential energy curves denoted by black and red colors, respectively. The zero slope of the potential energy in the relaxed system, reveals nearly full suppression of $E_{dep}$ in those bulk-like domain regions.}
\label{fig_S15}
\end{figure}

\newpage
\begin{figure}[ht]
\centering
\includegraphics[width=4.1in]{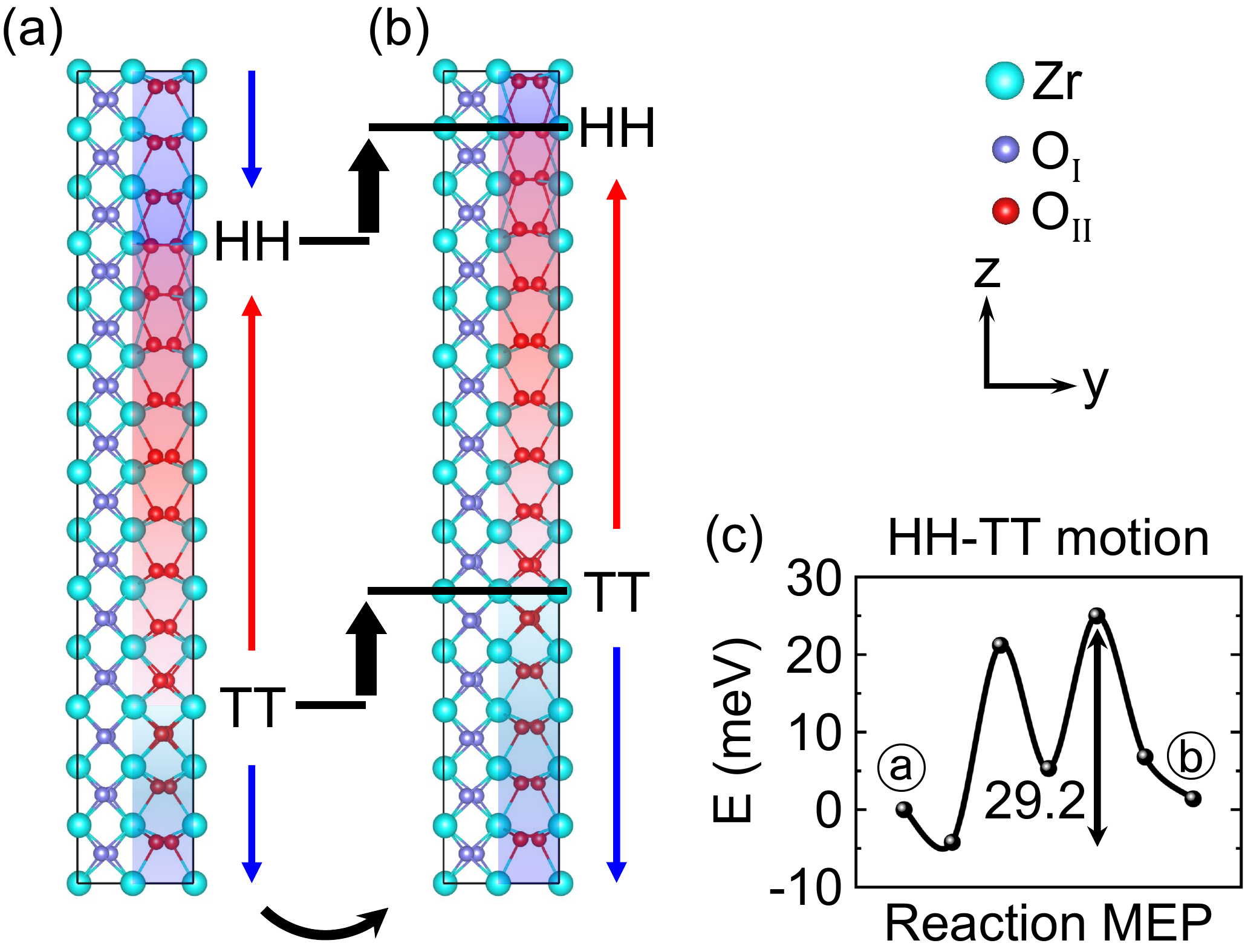}
\caption{\textbf{Properties of HH-TT collective wall motion in ZrO\textsubscript{2}} \textbf{(a,b)} Combined propagation of HH and TT CDWs and \textbf{(c)} the corresponding energy barrier (29.2 meV) and minimum-energy path (MEP) of the reaction.}
\label{fig_S16}
\end{figure}

%\end{document}

\end{document}